\documentclass[conference]{IEEEtran}
\ifCLASSINFOpdf
\else
\fi

\usepackage[linesnumbered,ruled,vlined]{algorithm2e}
\usepackage{algpseudocode}  
\usepackage{amsmath}  
\usepackage{threeparttable}%
\usepackage{multirow}%
\usepackage{booktabs}%
\usepackage{subfigure}%
\usepackage{textcomp}
\usepackage{color, balance}
\usepackage{array}
\usepackage{booktabs}
\usepackage{amsmath,amssymb,amsfonts}
\usepackage{marvosym}
\usepackage{colortbl}
\usepackage{wasysym}
\usepackage{utfsym}
\definecolor{mygray}{gray}{.9}
\usepackage{svg}
\usepackage{hyperref}
\newcommand{\paraspace}{\vspace{0.01in}}
\newcommand{\parab}[1]{\paraspace\noindent{\bf #1}}

\newcommand{\zq}[1]{\textcolor{black}{#1}}

\makeatletter
\newcommand{\rmnum}[1]{\romannumeral #1}
\newcommand{\Rmnum}[1]{\expandafter\@slowromancap\romannumeral #1@}
\makeatother
\newcommand{\eg}{\emph{e.g.,}\xspace}
\newcommand{\ie}{\emph{i.e.,}\xspace}
\newcommand{\et}{\emph{et al.}\xspace}
\newcommand{\ack}{\texttt{ACK}\xspace}
\newcommand{\sack}{\texttt{SACK-ACK}\xspace}
\newcommand{\synack}{\texttt{SYN/ACK}\xspace}
\newcommand{\syn}{\texttt{SYN}\xspace}
\newcommand{\rst}{\texttt{RST}\xspace}
\newcommand{\first}{\textsf{(i)}\xspace}
\newcommand{\second}{\textsf{(ii)}\xspace}

\newcommand{\tabincell}[2]{\begin{tabular}{@{}#1@{}}#2\end{tabular}}

\definecolor{fix_color}{rgb}{0,0,0}

\begin{document}
%
\title{Off-Path TCP Hijacking in Wi-Fi Networks: A Packet-Size Side Channel Attack}

	

%

\author{
\IEEEauthorblockN{Ziqiang Wang\IEEEauthorrefmark{1}\textsuperscript{\Letter},
Xuewei Feng\IEEEauthorrefmark{2},
Qi Li\IEEEauthorrefmark{2}\IEEEauthorrefmark{3}, 
Kun Sun\IEEEauthorrefmark{4}, Yuxiang Yang\IEEEauthorrefmark{2}, Mengyuan Li\IEEEauthorrefmark{5}, Ganqiu Du\IEEEauthorrefmark{6},\\ Ke Xu\IEEEauthorrefmark{2}\IEEEauthorrefmark{3}\textsuperscript{\Letter}  and Jianping Wu\IEEEauthorrefmark{2}\textsuperscript{\Letter}}

\IEEEauthorblockA{\IEEEauthorrefmark{1}Southeast University,
\IEEEauthorrefmark{2}Tsinghua University,
\IEEEauthorrefmark{3}Zhongguancun Lab,
\IEEEauthorrefmark{4}George Mason University,}

\IEEEauthorblockA{
\IEEEauthorrefmark{5}University of Toronto, \IEEEauthorrefmark{6}China Software Testing Center}

ziqiangwang@seu.edu.cn, fengxw06@126.com, \{qli01@, yangyx22@mails., xuke@\}tsinghua.edu.cn,\\  ksun3@gmu.edu, alyssamengyuanli@gmail.com, duganqiu@cstc.org.cn, jianping@cernet.edu.cn
}


\IEEEoverridecommandlockouts
\makeatletter\def\@IEEEpubidpullup{6.5\baselineskip}\makeatother
\IEEEpubid{\parbox{\columnwidth}{
		Network and Distributed System Security (NDSS) Symposium 2025\\
		23 - 28 February 2025, San Diego, CA, USA\\
		ISBN 979-8-9894372-8-3\\
		https://dx.doi.org/10.14722/ndss.2025.23305\\
		www.ndss-symposium.org
}
\hspace{\columnsep}\makebox[\columnwidth]{}}

\maketitle

\begin{abstract}
In this paper, we unveil a fundamental side channel in Wi-Fi networks, specifically the observable frame size, which can be exploited by attackers to conduct TCP hijacking attacks. 
Despite the various security mechanisms (\eg WEP and WPA2/WPA3) implemented to safeguard Wi-Fi networks, our study reveals that an off-path attacker can still extract sufficient information from the frame size side channel to hijack the victim's TCP connection. 
Our side channel attack is based on two significant findings: (i) response packets (\eg \ack and \rst) generated by TCP receivers vary in size, and (ii) the encrypted frames containing these response packets have consistent and distinguishable sizes.
By observing the size of the victim's encrypted frames, the attacker can detect and hijack the victim's TCP connections. 
We validate the effectiveness of this side channel attack through two case studies, \ie SSH DoS and web traffic manipulation.
Precisely, our attack can terminate the victim's SSH session in 19 seconds and inject malicious data into the victim's web traffic within 28 seconds.
Furthermore, we conduct extensive measurements to evaluate the impact of our attack on real-world Wi-Fi networks. We test 30 popular wireless routers from 9 well-known vendors, and none of these routers can protect victims from our attack. Besides, we implement our attack in 80 real-world Wi-Fi networks and successfully hijack the victim's TCP connections in 75 (93.75\%) evaluated Wi-Fi networks. 
We have responsibly disclosed the vulnerability to the Wi-Fi Alliance and proposed several mitigation strategies to address this issue.
\end{abstract}

\section{Introduction}
Nowadays, public Wi-Fi networks are widely available in various places, such as airports, coffee shops, hotels, and libraries.
Serving as a prevalent method for Internet access, Wi-Fi networks have undergone substantial advancements in security mechanisms, progressing from WEP to WPA3, to counter various crypto-cracking attacks~\cite{ohigashi2009practical, tews2009practical, vanhoef2017key, vanhoef2018release}. Consequently, it becomes difficult for an off-path attacker to get useful information (\eg the random sequence and acknowledgment numbers of TCP connections) from the encrypted Wi-Fi frames.
Additionally, certain security policies (\eg AP isolation and rogue AP detection~\cite{Linksys-rogueAP, Huawei-rogueAP}) are proposed to counteract ARP poisoning and rogue APs.
Moreover, recent efforts have rectified certain implementation vulnerabilities to thwart attackers from manipulating the router's transmission queues~\cite{schepersframing}, NAT mappings~\cite{yangexploiting}, and the next-hop routing via malicious ICMP redirects~\cite{feng2022man}. As a result, it poses a challenge for off-path attackers to hijack Wi-Fi network traffic.

However, in this paper, we demonstrate that the encrypted frame size constitutes a reliable side channel that can be exploited by attackers to conduct TCP hijacking attacks, even in Wi-Fi networks with AP isolation enabled. 
Precisely, we discover that TCP packets can be identified by analyzing the size of the encrypted wireless frames, thus allowing an attacker residing in the same Wi-Fi network to infer the state of the victim's TCP connection.
By exploiting this side channel (\ie the encrypted frame size), the attacker can infer the random sequence and acknowledgment numbers of the victim's TCP connection. Consequently, the attacker can pretend to be one peer of the victim's connection to either terminate the connection or inject malicious data into the connection, \ie hijacking the connection completely.

Our attack consists of four steps. First, the attacker accesses a public Wi-Fi network and probes alive supplicants in the WLAN. 
The attacker crafts ARP requests in the WLAN to identify alive supplicants\footnote{Attackers can also identify alive supplicants by exploiting the DHCP mechanism, especially to circumvent the AP isolation mechanism enabled in Wi-Fi networks. Refer to Section~\ref{sec:practical} for more details.}.
%
By collecting the ARP replies, the attacker can obtain the $<MAC, IP>$ address pair of each alive supplicant which is also a potential victim client of our TCP hijacking attack.
Then through analyzing the MAC address field of the captured wireless frames in the shared Wi-Fi channels, the attacker can filter the encrypted frames belonging to the victim client. If the Wi-Fi network provides multiple access channels, the attacker can scan all Wi-Fi channels to filter the victim's frames. Once the victim's frames are sniffed, the attacker gains a potent side channel to conduct the TCP hijacking attack.


%

Armed with this side channel (specifically, the victim's encrypted frame size), 
the attacker can detect TCP connections issued by the victim supplicant through manipulating the challenge \ack mechanism~\cite{ramaiah2010improving}. The attacker impersonates the victim supplicant and sends forged \synack packets to the server. 
If a TCP connection exists between the victim supplicant and the server, the server will reflect a TCP challenge \ack packet to the supplicant.
This challenge \ack packet (always encrypted as a 68-byte wireless frame at the link layer) will be sniffed by the attacker at the shared Wi-Fi channel.
By contrast, if no TCP connection exists, the attacker will not capture the 68-byte encrypted frame that carries the challenge \ack packet.
Based on this key difference, the attacker can easily detect a target TCP connection between the identified victim supplicant and a remote server.
Note that our attack does not directly exploit the vulnerability in the challenge \ack mechanism~\cite{cao2016off, DBLP:journals/ton/CaoQWDKM18}. Instead, we only use the challenge \ack mechanism as a trigger condition to assist our observations. 

%

Third, the attacker infers the sequence number of the target TCP connection.
%
The attacker pretends to be the victim supplicant and crafts TCP packets to the server. Those crafted TCP packets carry the guessed sequence numbers. 
If the guessed sequence number is less than the next sequence number to be received, the server will return a \ack packet carrying the SACK\footnote{Selective acknowledgment (SACK) is an option in TCP that allows a receiver to acknowledge non-contiguous blocks of data received from the sender. In this paper, we focus on exploiting the duplicate SACK option specified in RFC 2883~\cite{rfc2883}.} option in the TCP header to the supplicant. The SACK option in the TCP header will consume extra bits within the wireless frame.
In contrast, if the attacker specifies a sequence number greater than the next sequence number, the return \ack packet from the server will not carry the SACK option.
%
This subtle difference (\ie the variation in frame size) can be observed by the attacker to infer the correct sequence number. 
%

Fourth, the attacker proceeds to send forged \ack packets to the server, containing guessed acknowledgment numbers. If the specified acknowledgment number in the crafted TCP packet is below the server's accepted window,
the server will reflect a challenge \ack packet (68-byte encrypted frame) to the victim supplicant. Otherwise, the server will discard the crafted packet or accepted it silently. 
By analyzing the size of the victim's encrypted frames, the attacker can easily infer the acknowledgment number of a target TCP connection. At this stage, the attacker has gathered all the necessary elements to hijack a TCP connection.


We conduct a comprehensive measurement study to show that our attack can be performed to cause serious damage in the real world, \eg terminating a victim SSH connection or poisoning a web traffic within 28 seconds.
We test 30 popular wireless routers from 9 well-known vendors, and we discover that none of these routers can protect victims from our attack.
Besides, we evaluate our attack in 80 real-world Wi-Fi networks, including most popular Wi-Fi scenarios (\eg Wi-Fi networks in coffee shops, bookstores, enterprises, and restaurants). The experimental results show that 75 (93.75\%) out of the 80 evaluated Wi-Fi networks are vulnerable to our TCP hijacking attack.   

Finally, we have responsibly reported this vulnerability to the Wi-Fi Alliance and they have acknowledged the issue.
Currently, we are discussing the mitigation measures with the Wi-Fi Alliance.
The root cause of this vulnerability lies in the fixed size of Wi-Fi frames at the link layer, which inadvertently creates a reliable side channel and leaks information about TCP connections.
As a result, we propose two possible countermeasures: (\rmnum{1}) Modifying the 802.11 standards and dynamically padding the encrypted frames to prevent information leakage. (\rmnum{2}) Revisiting the TCP specifications so that the server responds consistently to different conditions.
The proof-of-concept (PoC) code for our attack is available at \href{https://github.com/Internet-Architecture-and-Security/Packet-Size-Side-Channel-Attack}{https://github.com/Internet-Architecture-and-Security/Packet-Size-Side-Channel-Attack}.

\textbf{Contributions.} Our main contributions are as follows:
\begin{itemize}

\item We uncover a fundamental side channel in Wi-Fi networks, \ie the observable frame size, which is inherent in all generations of Wi-Fi standards.
\item We show that this frame size side channel can be exploited by off-path attackers to infer the random sequence and acknowledgment numbers of TCP connections issued by victim clients in Wi-Fi networks, thus hijacking the target connections completely.


\item We conduct an extensive investigation against {\color{fix_color}30} popular AP routers and {\color{fix_color}80} real-world Wi-Fi networks. The experimental results show that our attack can cause serious damage in the real world.

\item We provide a thorough analysis on the root cause of the identified attack and discuss possible defenses to alleviate this attack. 

\end{itemize}


\noindent\textbf{Ethical Considerations.} When we evaluate the impact of our attack in the real world, we carefully design and conduct the following experiments to avoid causing damage or negative impacts on operational Wi-Fi networks.
Firstly, we provide a detailed explanation of our experimental procedure to the administrators and obtain their approval prior to conducting any tests.
Secondly, our testing does not affect other supplicants or compromise the capabilities of the Wi-Fi network. 
Precisely, in the SSH DoS attack, we take our laptop as the victim client and utilize our cloud server as the SSH server. 
In the web manipulation attack, the poisoned client is under our control (\ie our laptop), and the web server is not affected.
Third, we provide feedback to the administrators at the end of our experiments.

\section{Background}
\label{sec:background}

This section begins with an introduction to the 802.11 frame format and the security mechanisms in Wi-Fi networks. Following that, we briefly review the challenge \ack mechanism and the TCP options that can be used to facilitate our attack.

\begin{figure}[htbp]
\centerline{\includegraphics[width=\linewidth]{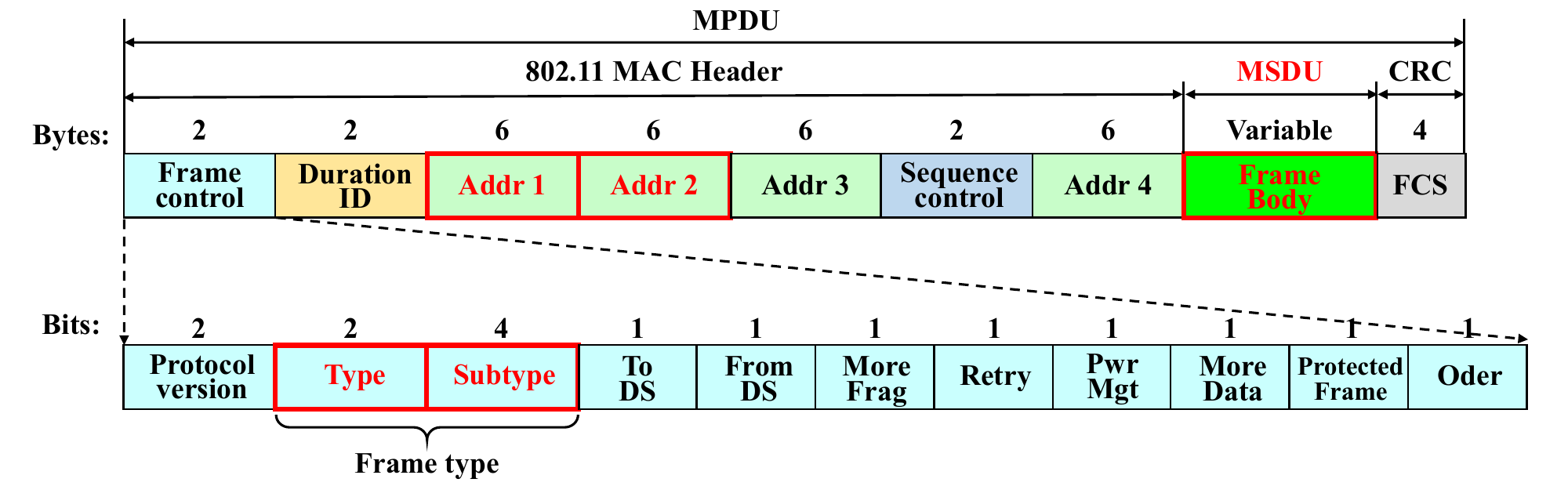}}
\caption{Layout of the 802.11 frame.}
\label{fig:1}
\end{figure}

\subsection{Frame Format and Security Mechanisms in Wi-Fi Network}

\parab{The 802.11 Frame Layout.}
Figure~\ref{fig:1} shows the layout of the 802.11 frame.
Firstly, the Frame Control (FC) field contains several flags and defines the type of the frame. The Type and Subtype fields together identify the function of the frame. There are currently three types (\ie management, control, and data frames) and more than 50 subtypes defined in 802.11 specifications. 
In our attack, the attacker needs to monitor the victim's TCP packets which will be encapsulated into 802.11 frames with type 2 and subtype 8 in Wi-Fi networks.
To identify the victim's encrypted frames, the attacker needs to analyze the addresses of the 802.11 frames.
There are four address fields in the 802.11 frame format. These fields are used to indicate the basic service set identifier (BSSID), source address (SA), destination address (DA), transmitting address (TA), and receiving address (RA). Certain frames might not contain some of the address fields. Certain address field usage is specified by the relative position of the address field (1 – 4) within the MAC header, independent of the type of address present in that field. Specifically, the Address 1 field always identifies the intended receiver(s) of the frame, and the Address 2 field, where present, always identifies the transmitter of the frame~\cite{9363693}.
In our attack, the attacker can identify the victim supplicant's encrypted frames through addresses 1 (RA) and address 2 (TA).
After filtering the victim's encrypted frames, the attacker needs to further analyze the payload size of the encrypted frames.
The payload (\ie MSDU in Figure~\ref{fig:1}) of a normal data frame contains the upper layer data (\eg TCP packets). The MSDU typically starts with an LLC/SNAP header and is protected by cryptographic encapsulation mechanisms (\ie TKIP, CCMP, and GCMP). In this paper, we refer to the encrypted frame size as the MSDU size.

\parab{Security Mechanisms in Wi-Fi Network.}
When connecting to a Wi-Fi network, the supplicant initiates a four-step handshake with the access point (AP) to establish a distinctive random session key\footnote{If the AP uses the outdated WEP encryption mechanism, there is no four-step handshake to negotiate the encryption key.}. Subsequently, both the supplicant and the AP utilize this session key to encrypt Wi-Fi frames and transmit them over the wireless channel~\cite{9363693}.
802.11i~\cite{1318903} outlines the requirements and procedures for ensuring the confidentiality of user information during wireless transmission, as well as the authentication of devices conforming to the IEEE 802.11 standard.
For an extended period, the security mechanisms employed by Wi-Fi networks (\eg WPA2 and WPA3) have primarily emphasized the improvement of confidentiality and data authentication. There has been a general belief that uncracked encrypted frames are secure.
However, in this paper we show that the encrypted frame size inadvertently forms a side channel which leaks information about the victim applicant in the Wi-Fi network.
It is worth noting that our attack does not sniff the four-step handshake frame to obtain the random session key. Instead, the attacker can directly exploit the size of encrypted frames within the Wi-Fi channel to launch a TCP hijacking attack.

\subsection{Challenge \ack Mechanism in TCP}
\label{sec:2.2}

\parab{Challenge \ack Mechanism.} 
The challenge \ack mechanism, proposed in RFC 5961~\cite{rfc5961}, serves as a defense against blind in-window attacks carried out by off-path attackers. 
In essence, the challenge \ack mechanism introduces more stringent requirements for TCP segment acceptance, where the receiver expects the sender to respond with the precise sequence number instead of falling within the receive window. This effectively thwarts blind injection attacks by off-path attackers. However, we demonstrate that this mechanism can be exploited to infer TCP connection information in the following manner.

Our attack leverages the trigger conditions of the challenge \ack mechanism in two distinct ways. Firstly, when a receiver detects an incoming \syn packet within an established TCP connection, regardless of the sequence number, it responds by sending an \ack (referred to as the challenge \ack) to the remote peer.
This \ack serves as a challenge for the remote peer to confirm the loss of the previous connection and the initiation of a new connection. Only the legitimate peer will receive this \ack and respond with a \rst segment containing the correct sequence number, derived from the \ack field of the challenge \ack packet, in the event of connection loss. Consequently, a spoofed \syn packet will generate an additional \ack, which will be disregarded by the peer as a duplicate \ack and will have no impact on the established connection. We will demonstrate how this challenging condition can be exploited to detect a victim TCP connection in Section~\ref{network_scanning}.

Secondly, the receiver employs a verification process for the acknowledgment number of each TCP segment to prevent blind data injection attacks.
Acceptance of an acknowledgment number for any data segment is contingent upon its falling within the range of ($SND.UNA - SND.WND, SND.NXT$), where $SND.UNA$ represents the sequence number of the first unacknowledged octet, $SND.WND$ denotes the maximum window size observed by the receiver from the sender, and $SND.NXT$ is the next sequence number to be sent, \zq{as illustrated for case (ii) in Figure~\ref{fig:ack}}.
If the acknowledgment number of the segment ($SEG.ACK$) is in the range ($SND.UNA - (2^{31} - 1), SND.UNA - SND.WND$), the receiver responds with a challenge \ack (see case (i) in Figure~\ref{fig:ack}). If the $SEG.ACK$ is greater than $SND.NXT$, the receiver silently discards this TCP segment. That can be exploited by attackers to infer the acceptable acknowledgment number, as described in Section~\ref{sec:ack}. 

\begin{table}[htbp]
\small
\centering
    \begin{threeparttable}
    \caption{TCP packet size analysis with IPv4.}
    \setlength{\tabcolsep}{1.2mm}
    \begin{tabular}{c | c  c | c c}
    \bottomrule
        \textbf{\multirow{2}{*}{\tabincell{c}{Packet \\type}}} & \multicolumn{2}{c|}{\textbf{TCP options}} & \textbf{\multirow{2}{*}{\tabincell{c}{Packet size\\ (Byte)}}} & \textbf{\multirow{2}{*}{\tabincell{c}{Frame size\\ (Byte)}}} \\
        \cline{2-3}
        & \tabincell{c}{\textbf{Timestamp}} & \textbf{SACK} & & \\
        \hline
        RST & \textbf{-} & \textbf{-} & 54 & 56\\
        ACK & \textbf{+} & \textbf{-} & 66 & 68\\
        SACK-ACK & \textbf{+} & \textbf{+} & 78 & 80\\
    \toprule
    \end{tabular}
    \begin{tablenotes} 
        \item \textbf{+} represents carrying the option, while \textbf{-} represents not carrying the option.
    \end{tablenotes} 
    \label{tab:packet_size}
    \end{threeparttable}
\end{table}

\subsection{TCP Options}

TCP options are supplemental fields that can be appended to the TCP header, offering added functionality and control. These options extend beyond the standard 20-byte TCP header and possess a variable size, not exceeding 40 bytes, contingent on the number of options included. 
Among the various TCP options available, the timestamp and selective acknowledgment options are the most commonly used.

\parab{Timestamp Option.} 
The TCP timestamp option is defined in RFC 1323~\cite{rfc1323}. It is widely used in modern operating systems and various studies~\cite{8526846, giffin2003covert, honda2011still}.
The timestamp option field spans a size of 10 octets, encompassing the timestamp value and timestamp echo reply fields. In practice, the timestamp option is typically padded with two extra bytes to maintain alignment of the TCP header on a 32-bit boundary.
In a TCP connection with timestamp functionality enabled, the \ack packet includes a timestamp value indicating its transmission time. This timestamp can be utilized by the sender to calculate round-trip time and estimate the current network state. However, \rst packets, which are employed for connection termination and lack TCP header options like the timestamp option, possess different sizes compared to \ack packets. This disparity in size between \rst and \ack packets is illustrated in Table~\ref{tab:packet_size}.
In this paper, we will show that the different size of the \ack packet and the \rst packet can be used to infer the source port number of a target TCP connection.

\parab{Selective Acknowledgment Option.}
%
\zq{The TCP selective acknowledgment (SACK) option is specified in RFC 2018~\cite{rfc2018} and extended in RFC 2883~\cite{rfc2883}.}
It is an optional feature that is typically enabled by default in the majority of TCP implementations. The SACK option is particularly recommended for networks experiencing frequent packet loss or packet reordering. Its utilization can significantly improve the performance and reliability of TCP connections in such environments.
\zq{In this paper, we mainly discuss the extended SACK option (also known as duplicate SACK) specified in RFC 2883. This extension to the SACK option allows the TCP sender to infer the order of packets received at the receiver, allowing the sender to infer when it has unnecessarily retransmitted a packet.}
When a receiver detects a TCP segment with a sequence number that has already been acknowledged as outdated, it responds by sending a \sack to notify the sender.
\zq{The sender could then use this information for more robust operations.}
%
However, if the sequence number of the received TCP segment has not yet been acknowledged, the receiver will reply with an \ack packet, which may have a different frame size (as shown in Table~\ref{tab:packet_size}) or may not respond at all, depending on the acknowledgment number of the segment.
In this paper, we will show that attackers in Wi-Fi networks can differentiate between these situations and thus infer the sequence number of a target TCP connection by analyzing the size of encrypted wireless frames.

\section{Threat Model}
\label{sec:threat_model}

\begin{figure}[htbp]
\centerline{\includegraphics[width=0.9\linewidth]{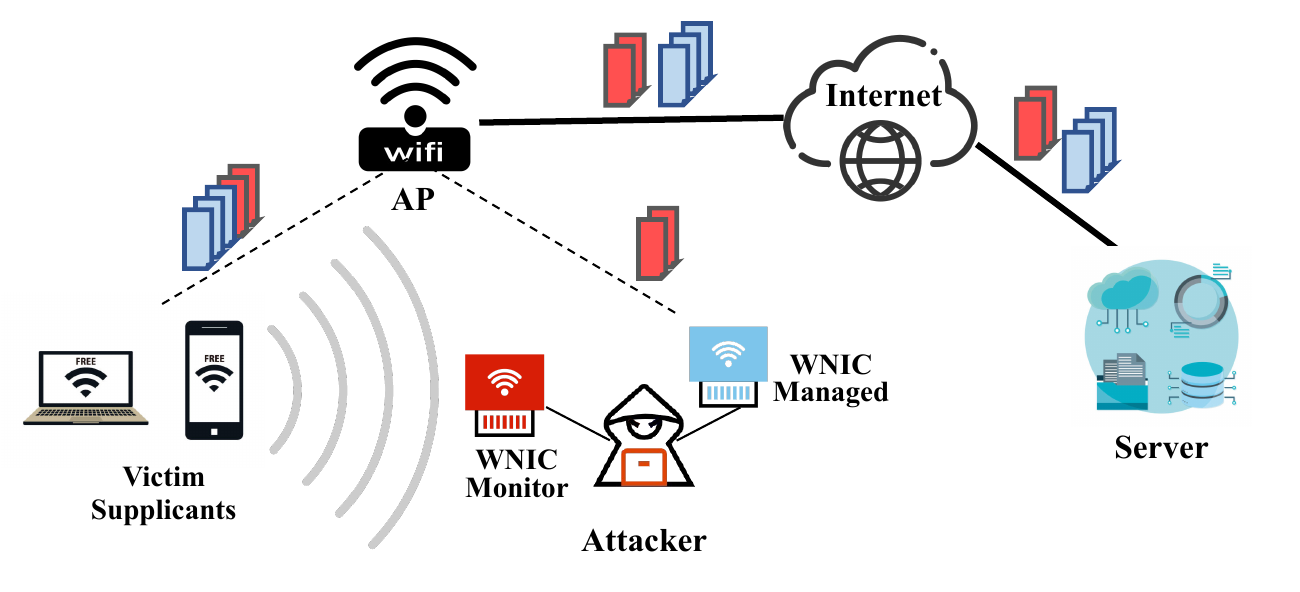}}
\caption{The threat model.}
\label{fig:2}
\end{figure}

Figure~\ref{fig:2} illustrates the threat model of our off-path TCP hijacking attack in Wi-Fi networks. The AP encrypts the network traffic of its supplicants via security mechanisms, \eg WPA2 or WPA3. A victim supplicant, such as a laptop or a smartphone, connects to the AP and establishes TCP connections with remote servers. The attacker, functioning as a regular supplicant without AP management privileges, utilizes multiple wireless network interface cards (WNICs). One (managed model) of these WNICs connects to the AP, while the others (monitor model) are utilized to sniff encrypted frames transmitted over the shared Wi-Fi channels. We make the assumption that the attacker has prior access to the target Wi-Fi network before performing our attack. This is a commonly accepted assumption in Wi-Fi hijacking scenarios, as highlighted in previous studies~\cite{feng2022man, vanhoef2021fragment, yangexploiting}. 

\begin{figure*}[t]
    \centering
    \begin{minipage}[t]{\linewidth}
        \subfigure[Identifying victim and detecting TCP connections.]{
            \begin{minipage}[t]
            {0.49\linewidth}
                \centering
                \includegraphics[width =\textwidth]{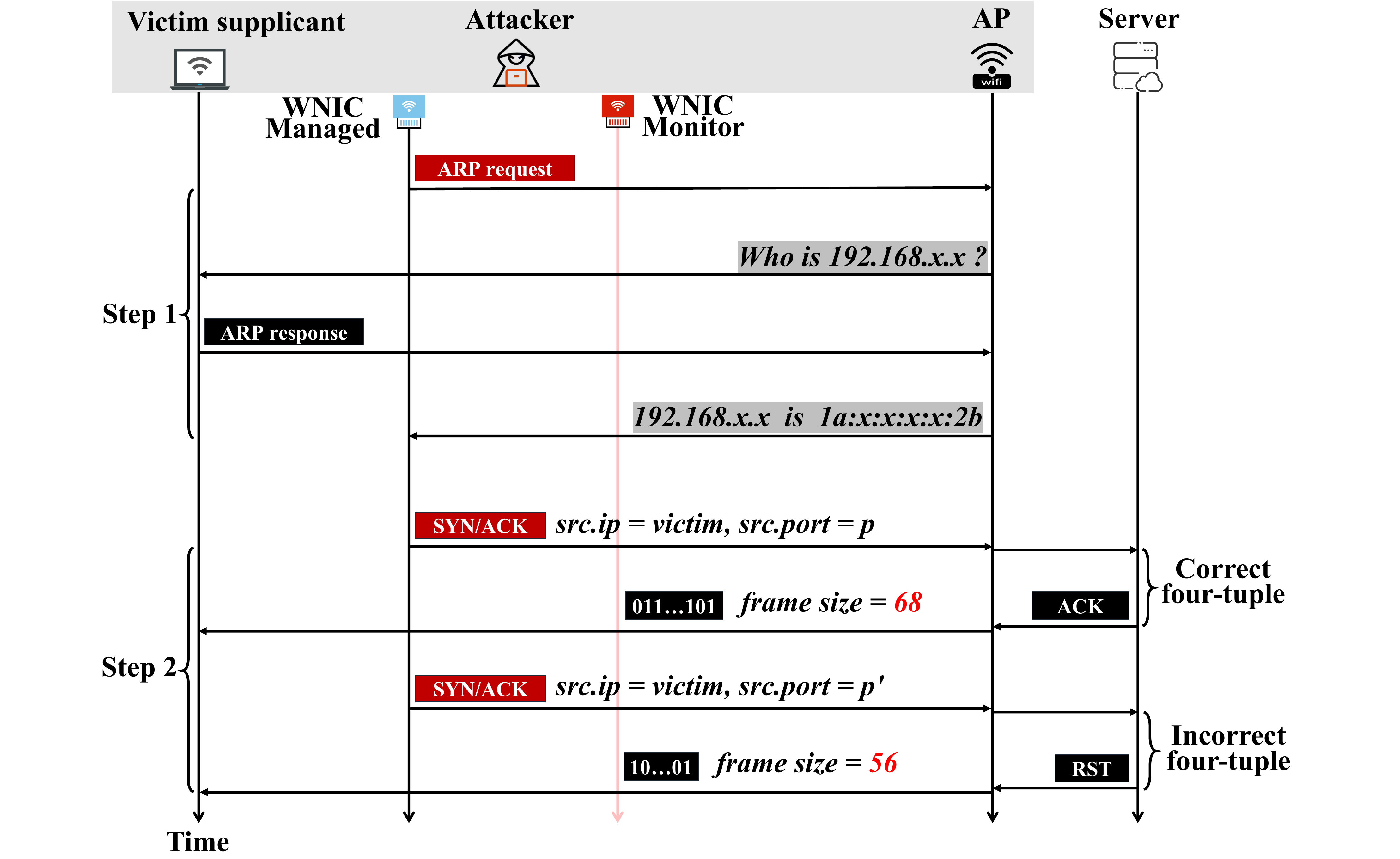}
                \label{attack_1_2}
            \end{minipage}}
        \subfigure[Inferring sequence and acknowledgment numbers.]{
            \begin{minipage}[t]
            {0.49\linewidth}
                \centering
                \includegraphics[width =\textwidth]{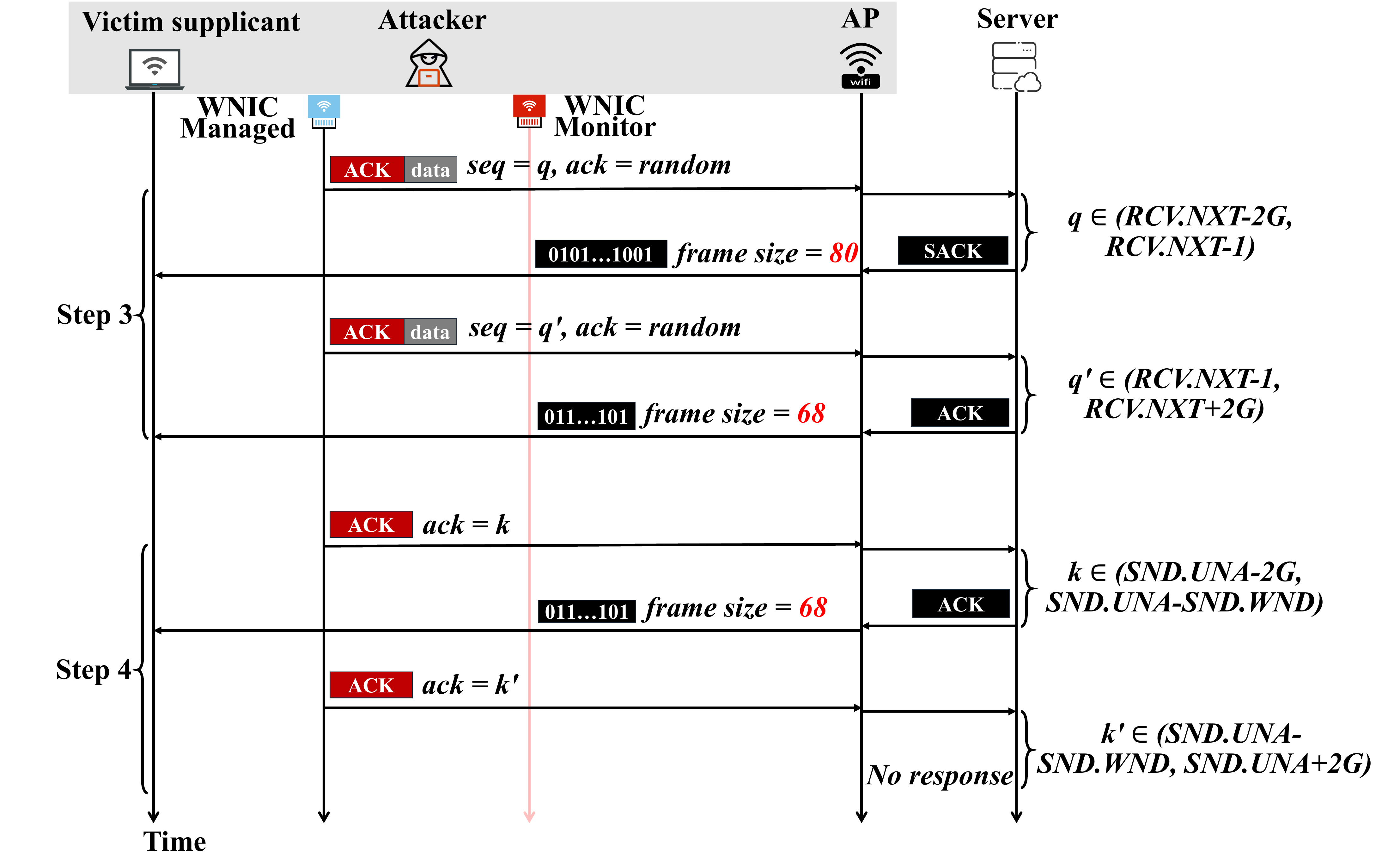}
                \label{attack_3_4}
            \end{minipage}}
    \end{minipage}
    \caption{Outline of our off-path TCP hijacking attack.}
    \label{fig:attack}
\end{figure*}

\section{TCP Hijacking with Encrypted Frame}
\label{sec:4}
Our attack exploits two key aspects. Firstly, the TCP stack exhibits inconsistent responses during packet verification. Depending on the validity of the received packet, the TCP receiver generates four different responses: no response packet, a \rst, an \ack, and a \sack. Due to the presence of TCP options, these responses can be distinguished based on their packet sizes. Secondly, the frames within the Wi-Fi network are observable, and the frame sizes of these responses are consistently fixed (see Table~\ref{tab:packet_size}). These characteristics create a significant side channel. An attacker can leverage this side channel to detect and hijack the victim's TCP connection.

\subsection{Attack Overview} 
Our TCP hijacking attack consists of four steps.

\noindent\textbf{Step 1: Identifying Victim.} The attacker accesses a Wi-Fi network and scans the WLAN for potential victim supplicants. In this step, the attacker identifies the $<MAC, IP>$ address pair of the victim to monitor its encrypted frames. 

\noindent\textbf{Step 2: Detecting TCP Connections.} After detecting potential victims alive in the WLAN, the attacker impersonates the victim supplicant\footnote{``Impersonating the victim supplicant" refers to the attacker specifying the source IP address of crafted packets as the victim’s IP address. This works in WLANs where the AP does not check the IP addresses.} and sends forged \syn/\ack packets to the server. At the same time, the attacker monitors the victim's encrypted frames in the Wi-Fi channel. By analyzing the encrypted frame size, the attacker can determine if a TCP connection exists between the victim and the server.

\noindent\textbf{Step 3: Inferring Sequence Number.} After detecting a victim's TCP connection, the attacker sends forged TCP packets 
with guessed sequence numbers to the server. These manipulated TCP packets prompt the server to generate \sack responses, which will be sniffed by the attacker when they (\ie 80-byte encrypted frames) are transmitted in the Wi-Fi channel. By monitoring the victim's encrypted frames, the attacker can identify the correct sequence number of the target TCP connection.

\noindent\textbf{Step 4: Inferring Acknowledgment Number.} With the inferred acceptable sequence number, the attacker proceeds to send forged \ack packets to the server. These \ack packets will trigger server's challenge \ack, which always appears as a 68-byte encrypted frame in the Wi-Fi network. By exploiting this challenge \ack, the attacker can locate the server's challenge \ack window and subsequently find an acceptable acknowledgment number.

After determining the sequence and acknowledgment numbers of the target TCP connection, the attacker can inject forged TCP packets into the connection with the intent to either terminate the connection or manipulate the data stream.

\subsection{Identifying Victim and Detecting TCP Connections}
\label{network_scanning}
\noindent\textbf{Identifying Victim.} The attacker first prepares the TCP hijacking attack from two aspects, \ie obtaining the $<MAC, IP>$ address pair of the victim and identifying the Wi-Fi channel used by the victim. The attacker actively sends ARP requests in the WLAN to detect other alive supplicants (\ie the potential victim clients of our TCP hijacking attack). By observing the ARP responses, the attacker can learn the victim's MAC address and IP address. 
With the victim's MAC address, the attacker sniffs encrypted frames in the Wi-Fi channel and filters the victim's frames based on address 1 (or address 2) in the 802.11 MAC header (see Figure~\ref{fig:1}). 
If the Wi-Fi network supports multiple accessed Wi-Fi channels, the attacker scans all Wi-Fi channels to identify the specific channel used by the victim. Subsequently, the attacker intercepts encrypted frames within the target Wi-Fi channel and filters out the victim's frames.

\noindent\textbf{Detecting TCP Connections.} 
With intercepting and analyzing the victim's encrypted frames, the attacker can identify the victim's TCP connections.
\zq{Typically, the attacker focus on detecting TCP connections between the victim and popular servers~\cite{tolley2021blind, cao2016off, yangexploiting}, such as servers of famous websites.}
A TCP connection is recognized by four elements, \ie [client IP address, client port number, server IP address, server port number]. 
\zq{The attacker can probe the server's IP address (e.g., using ``dig example.com") and access the server to determine the server's port number.}
%
Subsequently, the attacker needs to infer the client's IP address and port number. In our attack, the client IP is obtained via ARP response.
Thus, the last remaining element to determine is the client port number.

Given that a TCP connection was previously established by the legitimate user on a victim client using a source port \emph{p}, the attacker impersonates as the client and sends forged \synack packets to the server. As per the challenge \ack mechanism described in RFC 5961~\cite{rfc5961}, if the forged \synack packet contains the same client port number \emph{p}, the server will respond with a challenge \ack to the client. This challenge \ack packet will be encapsulated into a 68-byte encrypted frame and sniffed by the attacker, during transmission from the AP to the client.

In contrast, when the client port number specified in the forged \synack packet is not equal to \emph{p}, the server will reply with a \rst packet. This \rst packet is encapsulated within a 56-byte encrypted frame. Therefore, by examining the size of the encrypted frame, as depicted in step 2 of Figure~\ref{attack_1_2}, the attacker can determine whether the guessed client port number is correct or not.

The attacker iterates through the above procedure by changing the client port number specified in the forged \synack packet. This procedure continues until the correct port number \emph{p} is identified. Finally, the attacker identifies a target TCP connection operating on the four-tuple, \ie [client IP address, client port number, server IP address, server port number].

\subsection{Inferring Sequence and Acknowledgment Numbers}
\label{seq}

In this section, we begin with a concise overview of the mechanism used to verify the sequence number and acknowledgment number of TCP segments.
Next, we introduce the approach for inferring the exact sequence number and an acceptable acknowledgment number by leveraging encrypted Wi-Fi frames. 

\subsubsection{Verifying TCP Segment}
According to RFC 9293~\cite{rfc9293}, upon receiving a TCP segment, the TCP receiver first performs a verification by comparing the sequence number ($SEG.SEQ$) specified in the TCP header with its receive window. In other words, the condition $RCV.NXT \leq SEG.SEQ \leq RCV.NXT + RCV.WND$ must be met, where
$RCV.NXT$ denotes the next expected sequence number for an incoming segment, and $RCV.WND$ indicates the size of the receive window.
Furthermore, as per the specification, the \ack flag is consistently set to true, except for the initial \syn packet used for connection establishment. If the \ack bit is disabled, the receiver will discard the segment. Therefore, when hijacking the target TCP connection, the attack must infer an acceptable acknowledgment number and sequence number.

In practice, TCP operates in full duplex mode, thus allowing the attacker to infer the sequence and acknowledgment numbers in either direction. For instance, the client's $RCV.NXT$ (next expected sequence number) and $SND.NXT$ (next sequence number to be sent) are equivalent to the server's $SND.NXT$ and $RCV.NXT$~\cite{rfc9293}. In our attack, our main focus is on inferring the sequence and acknowledgment numbers that are deemed acceptable by the server side.

\subsubsection{Inferring the Exact Sequence Number}

\begin{figure}[tbp]
\centerline{\includegraphics[width=0.85\linewidth]{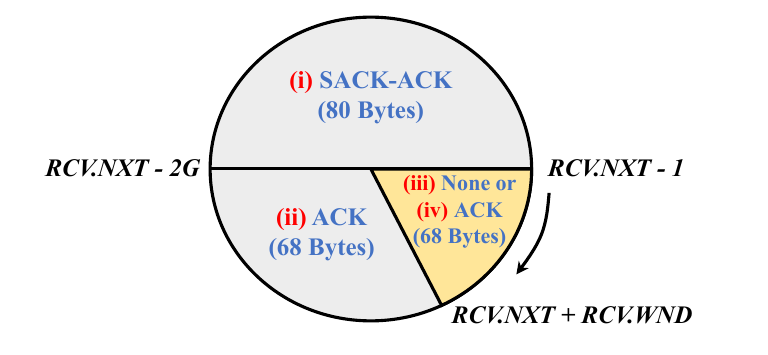}}
\caption{Sequence number window illustration.}
\label{fig:seq}
\end{figure}

To infer the exact sequence number on the server side, the attacker impersonates the client (\ie a victim supplicant in WLAN) and sends forged TCP packets containing data to the server. These packets carry the guessed sequence numbers and a random acknowledgment number. 
The sequence number space is $2^{32}$ (\ie 4G), and the server exhibits four distinct responses corresponding to different sequence numbers\footnote{The forged TCP packets with random acknowledgment numbers only elicit the server's response with the duplicate SACK option.}, as illustrated in Figure~\ref{fig:seq}.
(\rmnum{1}) If the guessed sequence number falls within the range of ($RCV.NXT - 2G, RCV.NXT - 1$), the server returns a \sack response with an encrypted frame size of 80 bytes. (\rmnum{2}) If the guessed sequence number exceeds the upper boundary of the acceptable window (\textit{i.e.,} $RCV.NXT + RCV.WND$), the server sends an \ack response consisting of a 68-byte encrypted frame to the client. (\rmnum{3}) If the guessed sequence number is deemed acceptable but the random acknowledgment number ($SEQ.ACK$) is invalid (\textit{i.e.,} $SEQ.ACK > SND.NXT$), the server silently discards the packet. (\rmnum{4}) If the guessed sequence number is deemed acceptable and the random acknowledgment number falls within the challenge window, the server responds with a challenge \ack in compliance with RFC 5961~\cite{rfc5961}.

The attacker's goal is to observe the \sack response, which is contained in an 80-byte encrypted frame. By examining the presence or absence of the \sack, as illustrated in step 3 of Figure~\ref{attack_3_4}, the attackers can determine if the guessed sequence number is less than $RCV.NXT - 1$ or greater than $RCV.NXT - 1$ (\ie identifying the exact sequence number). Employing a binary search strategy, the attacker can progressively refine their guesses and accurately identify the exact sequence number by analyzing the observed \sack responses.

\subsubsection{Inferring an Acceptable Acknowledgment Number}
\label{sec:ack}
To infer an acceptable acknowledgment number, the attacker firstly leverages the challenge \ack mechanism to locate the lower boundary of the challenge window. Then the attacker can easily obtain an acceptable acknowledgment number by adding $2^{31}$ (\ie 2G) to the lower boundary. 

\begin{figure}[htbp]
\centerline{\includegraphics[width=0.85\linewidth]{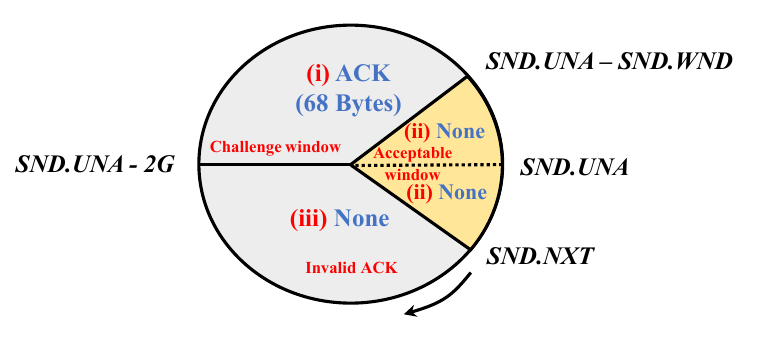}}
\caption{Acknowledgment number window illustration.}
\label{fig:ack}
\end{figure}

The challenge window for TCP segment acknowledgment number is defined in RFC 5961~\cite{rfc5961} (see Section~\ref{sec:2.2}). 
As outlined in RFC 5961, the acknowledgment number space can be divided into three distinct cases, as shown in Figure~\ref{fig:ack}. (\rmnum{1}) The acknowledgment number falls within the challenge window, defined as ($SND.UNA - 2G, SND.UNA - SND.WND$). (\rmnum{2}) The acknowledgment number resides within the acceptable \ack window, encompassing ($SND.UNA - SND.WND, SND.NXT$). (\rmnum{3}) Invalid acknowledgment numbers are those that exceed $SND.NXT$, denoted as $SEG.ACK > SND.NXT$.

In the first case (\ie falling within the challenge window), the receiver will respond with a challenge \ack to verify the legitimacy of the segment. In the second case, the receiver accepts the segment directly for further processing. Otherwise, the receiver will silently discard the TCP segment.
For an off-path attacker, the last two cases are indistinguishable. However, the attacker can determine the first case, where a 68-byte encrypted frame is observed.

To locate the server's challenge \ack window, the attacker impersonates the client and sends forged \ack packets to the server. The forged \ack packets carry the guessed acknowledgment number, as well as a sequence number in the server's acceptable window inferred in the previous step. 
If the attacker sniffs a returned 68-byte encrypted frames in the Wi-Fi channel, it indicates that the guessed acknowledgment number falls within the receiver's challenge window (as shown in step 4 of Figure~\ref{attack_3_4}). 
Typically, the window size of challenge \ack is between $2^{30}$ and $2^{31}$, \ie the challenge window is a quarter of the entire acknowledgment number range. Hence, the attacker can divide the acknowledgment range into four blocks and try at most four times to find an acknowledgment number ($ack\_challenge$) that is located in the challenge window.

After locating the server's challenge \ack window, the attacker can detect the lower boundary of the challenge \ack window.
In the beginning, the attacker locates the lower boundary of the range ($ack\_challenge-2G$, $ack\_challenge$). 
Subsequently, the attacker employs a binary strategy to progressively narrow down the detection range, ultimately determining the lower boundary of the challenge \ack window.
Once the lower boundary is detected, the attacker can get the server's $SND.UNA$ value by adding $2G$ to the lower boundary.
When all previously sent data has been acknowledged, the value of $SND.UNA$ is equal to $SND.NXT$.

\subsection{Practical Considerations}
\label{sec:practical}

\begin{figure}[htbp]
\centerline{\includegraphics[width=0.9\linewidth]{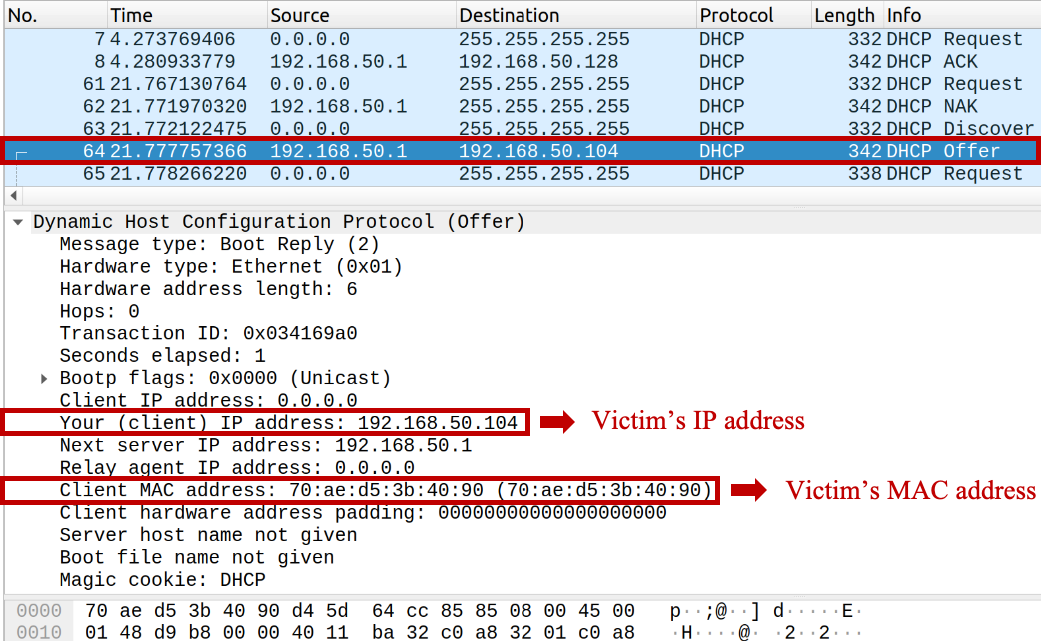}}
\caption{Snapshot of obtaining the victim's IP address via DHCP.}
\label{fig:dhcp}
\end{figure}

\parab{AP Isolation.} 
It also known as client isolation, is a security policy that can be implemented in wireless networks to separate individual devices or users from each other, enhancing network security and privacy.
In the Wi-Fi network with AP isolation enabled, the AP will discard ARP requests within the WLAN, preventing the attacker from obtaining the $<MAC, IP>$ address pair of the alive supplicant. 
In this case, the attacker can spoof the victim's MAC address and leverage the DHCP mechanism to obtain the victim's $<MAC, IP>$ address pair. Specifically, the attacker first sniffs the encrypted Wi-Fi frames and identifies the MAC address of the alive supplicant. Second, the attacker spoofs the victim's MAC address to authenticate with the AP and requests to lease a private IP address. As the DHCP server guarantees not to reallocate the leased address within the requested time and attempts to return the same network address each time the client requests an address~\cite{rfc2131}, the attacker will be assigned the same private IP address that the victim is leasing. Consequently, the attacker obtains the victim's $<MAC, IP>$ address pair (as shown in Figure~\ref{fig:dhcp}). If the victim uses Management Frame Protection (MFP), the attacker may encounter difficulties with AP authentication when spoofing the victim's MAC address. However, prior work has shown that implementation vulnerabilities can be abused to circumvent MFP~\cite{DBLP:conf/wisec/SchepersRV22, schepersframing}.

Note that our attack does not require overwriting the victim's security context to intercept the victim's packets~\cite{schepersframing}, but rather to obtain the victim's $<MAC, IP>$ address pair. Therefore, this approach does not necessitate the AP to support pairwise master key (PMK) caching for rapid connection.
%
With the victim's MAC and IP address in hand, the attacker proceeds to send forged packets to the server and detect the victim's TCP connection, as previously described. Armed with the inferred TCP connection information, the attacker can terminate or manipulate the target TCP connection. 
Due to AP isolation, the attacker is unable to send packets directly to the victim supplicant. Nevertheless, the attacker can opt to send malicious packets to the AP's external IP address, which can be obtained through ICMP ping messages, as demonstrated in previous research~\cite{yangexploiting}.
These packets will be forwarded to the victim supplicant through the AP.


\parab{Background Traffic.} 
The background traffic may degrade the quality of the side channel (\ie victim's frame size), thereby impacting the effectiveness of the attack. 
Specifically, if the TCP packets in the background traffic have the same size as the response (\eg challenge \ack) from the victim server, the attacker may mistakenly identify them as the actual responses. 
During practical attacks, the server may send empty \ack packets (such as keep-alive \texttt{ACKs}) to the victim supplicant. 
These empty \ack will interfere with the attacker's ability to infer the port number and acknowledgment number of the TCP connection, as they share the same size as challenge \ack packets.
Fortunately, the attacker has the option to leverage \sack to complete the attack, thereby bypassing the need to contend with empty \ack packets.
%
Specifically, \first when inferring the port number, the attacker sends two TCP packets containing data to the server, each bearing sequence numbers $seq$ and $seq+2^{31}$ respectively. If the TCP port is accurately inferred, the attacker will encounter an 80-byte encrypted frame, as one of the two packets in question is bound to elicit the server's \sack response. Conversely, if the inference is incorrect, the attacker will not observe the 80-byte encrypted frame.
\second Since TCP is full duplex, the attacker can utilize \sack to infer the sequence number on the client side and consequently obtain the acknowledgment number on the server side.

\parab{Shifting Receive Window.} When the victim's TCP connection carries on ongoing traffic, the acceptable sequence and acknowledgment windows will shift during the attack. Fortunately, the attack can proceed as long as the inferred sequence number and acknowledgment number fall within the sliding window. The attacker can repeatedly infer the sequence number and acknowledgment number.
Even if the receive window slides quickly enough to thwart the attacker's inference, the attacker can opt to target the other end of the TCP connection.
\zq{In typical high-traffic scenarios like file downloading, the server-side sequence number triggers fast client-side acknowledgment, while the client-side sequence number grows slower. Attackers can infer the client-side sequence number and conduct brute-force attacks by sending multiple spoofed packets with acknowledgment numbers at different intervals, exploiting the large accepted window for acknowledgment number~\cite{DBLP:conf/TCP_spoof} at this stage.}

\section{Case Study Attacks}
\label{sec:imp}
In this section, we demonstrate two cases, \ie SSH DoS and web manipulation, to illustrate how TCP connections can be hijacked by exploiting the encrypted frame size in Wi-Fi networks.
In summary, an off-path attacker can reset an SSH service within 19 seconds and inject malicious data into a HTTP web page\footnote{HTTPS can prevent attackers from injecting malicious data. However, reports on HTTPS adoption~\cite{W3Techs} indicate that there are 15\% of websites still based on HTTP as of April 2024. Additionally, our measurements of the top 1 million websites based on Tranco~\cite{DBLP:conf/ndss/PochatGTKJ19} show that about 10\% of them based on HTTP.} within 28 seconds.

\subsection{TCP DoS Attack}
\label{DoS}

In this case, we demonstrate that an off-path attacker can reset the TCP connection between a victim client and a remote server, resulting in a DoS attack. We specifically conduct the attack under the common scenario of SSH.

\noindent\textbf{Experimental Setup.}
This case involves three hosts: an SSH server (a rented VPS) 
running OpenSSH 8.4 and OpenSSL 1.1.1, a victim client (our laptop) running MacOS, and an attacker equipped with Kali 2023.1 and multiple wireless network interface cards. 
The victim client is a supplicant in our Wi-Fi network and connects to the remote SSH server. The client will send commands to the server intermittently.
Note that although the attacker and the victim supplicant are in the same Wi-Fi network, the attacker does not know the session key between the victim supplicant and the AP. The attacker attempts to terminate the connection by impersonating the victim supplicant and sending forged \rst packets to the server. 
Taking into account the potential impact of Linux kernel versions, we strategically deploy servers with a variety of Linux kernel versions. Detailed configuration information for these servers is provided in Table~\ref{tab:dos}.

\noindent\textbf{Attack Procedure.}
In this attack, the off-path attacker needs to infer the 4-tuple of [client IP address, client port number, server IP address, server port number] and the exact sequence number of the target TCP connection. 
The server's IP address and port number are publicly known to the attacker~\cite{tolley2021blind, cao2016off, yangexploiting}, thus it only needs to identify the other three remaining elements to proceed with the attack. 
The attacker first probes the victim client's IP address and MAC address. 
Second, the attacker exploits the TCP header options to determine the client port number and infer the exact sequence number, as outlined in Section~\ref{sec:4}. Third, a crafted \rst packet carrying the inferred value is issued to the server, and the server will be tricked into terminating the current SSH connection with the victim client.

\begin{table}[htbp]\small
\centering
    \begin{threeparttable}
    \caption{Experimental results of SSH connection reset.}
    \begin{tabular}{c c c c c c}
    \toprule  
        \textbf{\tabincell{c}{Server \\address}}  & \textbf{\tabincell{c}{Linux \\version}} & \textbf{\tabincell{c}{Time \\cost (s)}} & \textbf{\tabincell{c}{Bandwidth \\cost (KB/s)}} & \textbf{\tabincell{c}{Success \\rate}} \\
        \midrule
        \rowcolor{mygray}
        82.x.x.41 & 5.4 & 18.47 & 77.04 & 8/10\\
        
        150.x.x.186 & 5.15 & 19.56 & 80.91 & 9/10\\
        \rowcolor{mygray}
        43.x.x.151 & 5.10 & 18.24 & 69.15 & 8/10\\
        
        43.x.x.84 & 4.15 & 17.26 & 68.18 & 8/10\\
        \rowcolor{mygray}
        43.x.x.187 & 3.13 & 20.12 & 82.07 & 9/10\\
    \bottomrule
    \end{tabular}
    \label{tab:dos}
    \end{threeparttable}
\end{table}

\begin{figure}[htbp]
\centerline{\includegraphics[width=\linewidth]{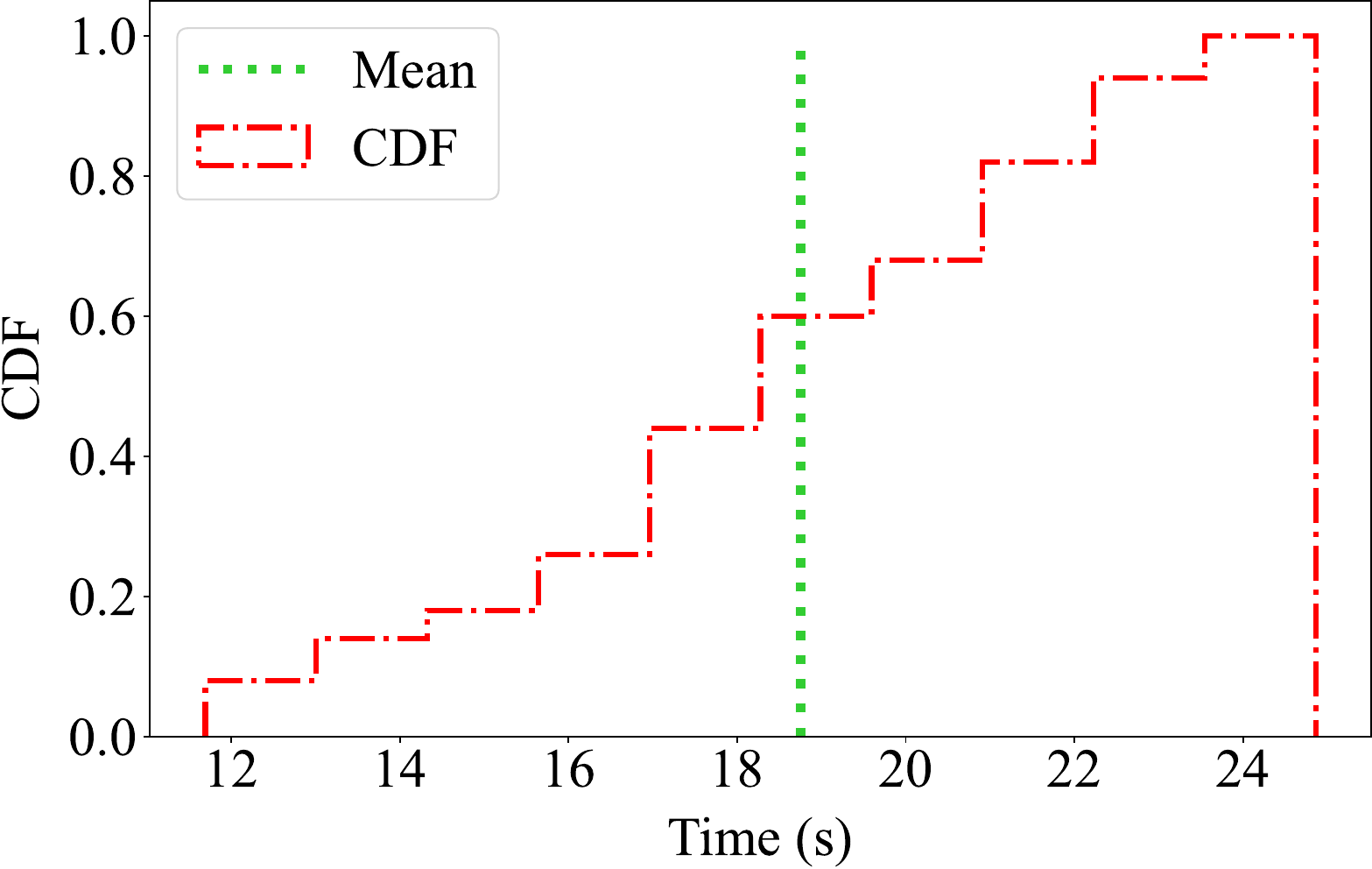}}
\caption{Empirical CDF of time cost of SSH connection reset.}
\label{fig:dos}
\end{figure}

\noindent\textbf{Results Evaluation.}
Table~\ref{tab:dos} displays the outcomes of our experiments, revealing that our attack is effective for different Linux versions.
In particular, our attack exhibits an average bandwidth consumption of 75.76 KB/s, while maintaining an average execution time of 18.78 seconds. The empirical time cost distribution is shown in Figure~\ref{fig:dos}.
Our attack achieves a success rate of 84\% on average. For the unsuccessful attempts, the primary cause is wireless interference,  leading to the attacker missing crucial encrypted frames belonging to the victim. These frames contain the server's responses to the probe packets. We will discuss wireless interference in depth in Section~\ref{sec:discussion}.

\subsection{TCP Manipulation Attack}
\label{web}
TCP connection hijacking poses a substantial threat to higher-layer applications, enabling malicious activities such as injecting harmful data into HTTP websites. 
As a case in point, we demonstrate that in a typical financial website scenario, an off-path attacker can hijack the underlying TCP connection, thereby tampering with real-time financial data displayed on the victim's web page.

\begin{figure}[tbp]
    \centering
    \begin{minipage}[t]{\linewidth}
        \subfigure[The attacker injects malicious data into the victim's web connection.]{
            \begin{minipage}[t]
            {\linewidth}
                \centering
                \includegraphics[width =0.9\textwidth]{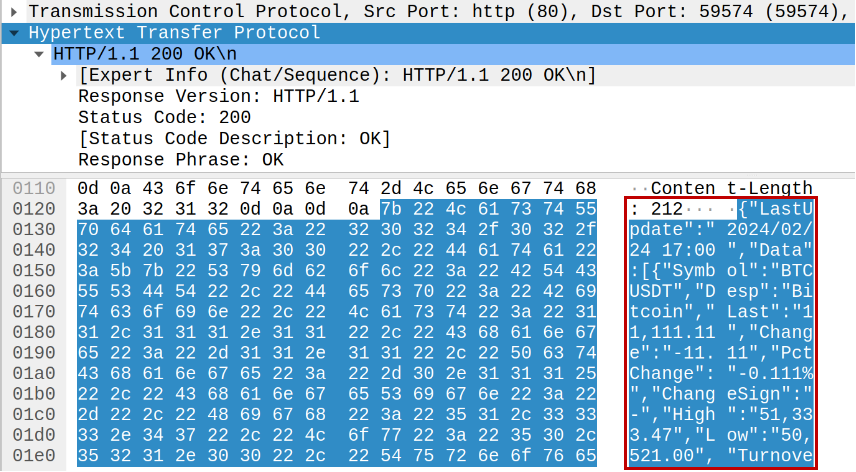}
            \label{fig:web_packet}
            \end{minipage}}
        \subfigure[The attacker manipulates the Bitcoin price presented on the victim's web page.]{
            \begin{minipage}[t]
            {\linewidth}
                \centering
                \includegraphics[width =0.9\textwidth]{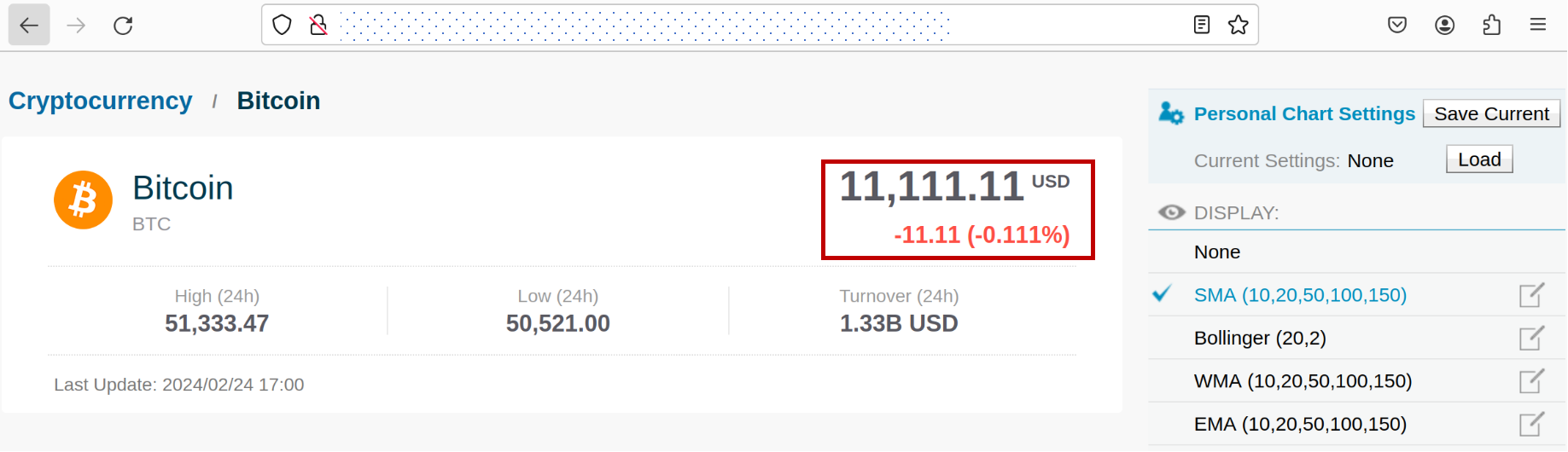}
                \label{fig:web}
            \end{minipage}}
    \end{minipage}
    \caption{Snapshots of web injection.}
    \label{fig:TCP hijack}
\end{figure}

\noindent\textbf{Experimental Setup.}
This attack involves three hosts: a web server, a victim client (our laptop), and an off-path attacker. We use a real financial website\footnote{For ethical considerations, we anonymize this financial website in the paper. Moreover, our attack do not affect the website, since we only manipulate the web cache of the client side, \ie our controlled laptop.} as the web server. 
This website employs HTTP to deliver real-time Bitcoin price to the client in JSON format at 5-second intervals. 
\zq{Before launching the attack, the attacker uses the dig tool to probe the server's IP address, explores the website to determine the server's port, and familiarizes themselves with the JSON data structure.}
%
%
Consequently, the attacker can determine the server's IP address and port, as well as identify specific data within the packet, as depicted in Figure~\ref{fig:web_packet}, enabling manipulation of the victim's web page.
The victim client browses financial information on the web page via Wi-Fi. 
Both the attacker and the victim client are connected to the same Wi-Fi network.
The off-path attacker attempts to detect and hijack the TCP connection between the victim client and the server. 
The server maintains a single long-lived TCP connection\footnote{As recommended in RFC 2616~\cite{rfc2616}, 
the client typically does not maintain multiple long-lived TCP connections with the server simultaneously.} with the client to transmit real-time financial data. 
Note that the modern browser may open multiple concurrent TCP connections along with the long TCP connection to speed up the page loading. These concurrent TCP connections are short-lived and have minimal influence on inferring the target long-lived TCP connection. Even if the server maintains multiple long-lived TCP connections with the client in some cases, the attacker can infer all the TCP connections and inject malicious data.

\noindent\textbf{Attack Procedure.}
The web connection hijacking attack consists of five steps: (\rmnum{1}) The attacker determines the MAC address and IP address of the victim client in the WLAN. (\rmnum{2}) By exploiting the encrypted frames, the attacker detects the client's port number to obtain the TCP 4-tuple information. (\rmnum{3}) The attacker infers the exact sequence number and (\rmnum{4}) gets an acceptable acknowledgment number. (\rmnum{5}) The attacker impersonates the server and injects forged TCP packets  with the inferred values into the victim client. 
Finally, the client will accept the forged TCP packets, which subsequently update the financial information on the web page.

\begin{figure}[htbp]
\centerline{\includegraphics[width=\linewidth]{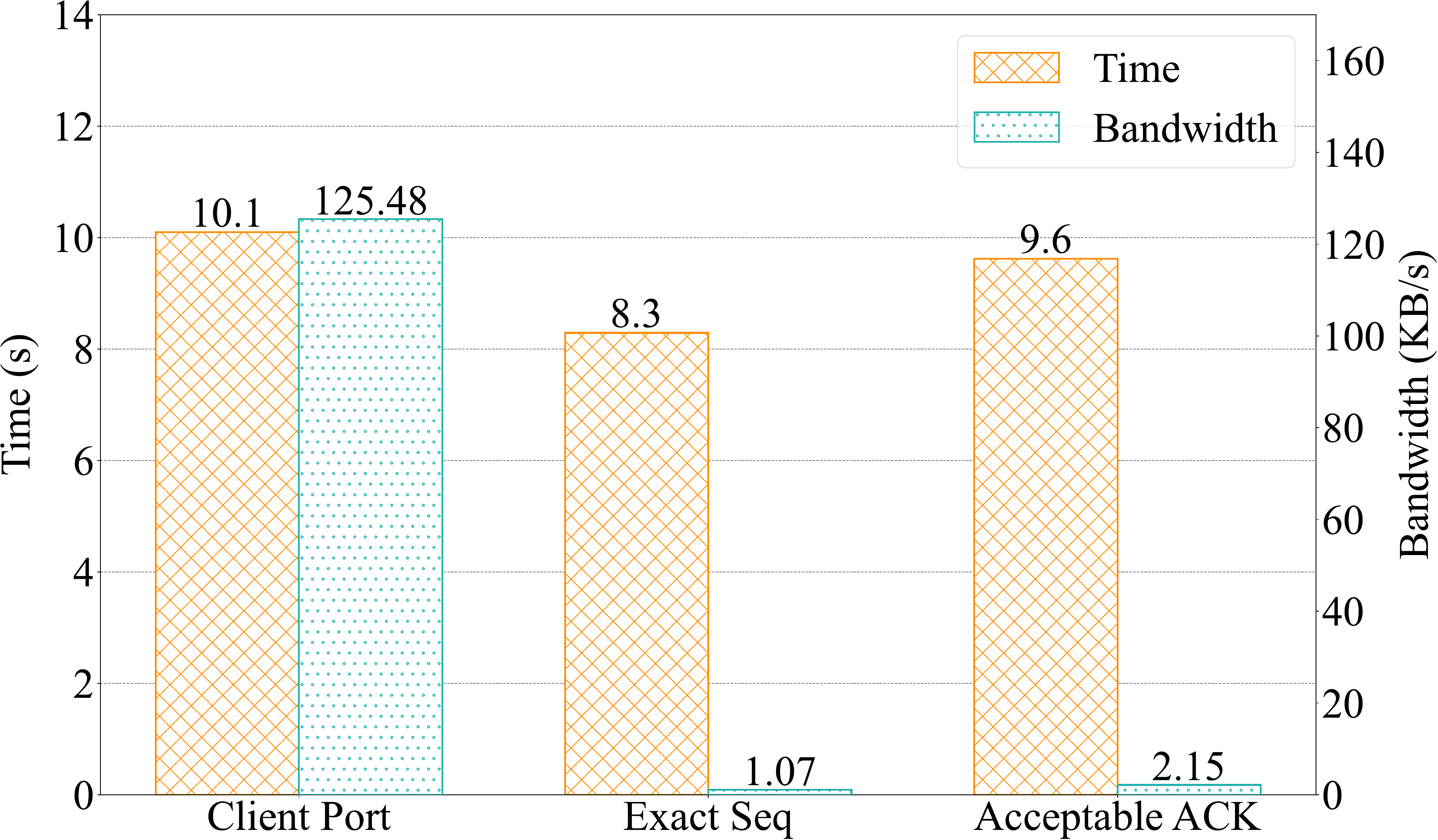}}
\caption{Time/Bandwidth overheads of web manipulation.}
\label{fig:web_cost}
\end{figure}

\noindent\textbf{Results Evaluation.} Figure~\ref{fig:web_cost} illustrates the time cost and bandwidth consumption during the attack. It takes an average of 10.1 seconds to identify the client port number 
and 8.3 seconds to find the exact sequence number. 
Time cost required to find an acceptable acknowledgment number takes 9.6 seconds.
The average duration of the entire attack is 28 seconds, with an average bandwidth cost of 46.32 KB/s. 
In this case, the attacker needs to infer an acceptable acknowledgment number, hence the success rate of this attack is lower than the TCP DoS attack but still exceeds 70\%.
After obtaining all the necessary information, the attacker sends forged TCP packets to the victim client and manipulates sensitive data on the web page. Figure~\ref{fig:web} shows a snapshot of the manipulated web page and where attacker alters the Bitcoin price.

\renewcommand{\dblfloatpagefraction}{.9}

\begin{table*}[htbp]\small
\centering
    \begin{threeparttable}
    \caption{Details of 30 tested wireless routers.}
    \begin{tabular}{c c c c c c c c}
    \toprule  
        \textbf{\tabincell{c}{Router}} & 
        \textbf{\tabincell{c}{Generation}} & \textbf{\tabincell{c}{WPA}} &
        \textbf{\tabincell{c}{IPv6 Enabled}} &
        \textbf{\tabincell{c}{Vendor}} &  \textbf{\tabincell{c}{Built-in Firewall}} & \textbf{\tabincell{c}{Anti-Flooding}} & \textbf{\tabincell{c}{MAC-ADDR \\Filtering}} 
        \\
        \midrule
        \rowcolor{mygray}
        Mi 4C & Wi-Fi 4 & WPA2 & No & Xiaomi & \CIRCLE & \CIRCLE & \CIRCLE
        \\
        \rowcolor{mygray}
        Redmi AC2100 & Wi-Fi 5 & WPA2 & Yes & Xiaomi & \CIRCLE & \CIRCLE & \CIRCLE
        \\
        \rowcolor{mygray}
        AX6000 & Wi-Fi 6 & WPA2/WPA3 & Yes & Xiaomi & \CIRCLE & \CIRCLE & \CIRCLE
        \\
        \rowcolor{mygray}
        AX9000 & Wi-Fi 6 & WPA2/WPA3 & Yes & Xiaomi & \CIRCLE & \CIRCLE & \CIRCLE
        \\
        \midrule
        TL-WR841N & Wi-Fi 4 & WPA2 & No & TP-LINK & \CIRCLE & \Circle & \CIRCLE 
        \\
        Archer AXE300 & Wi-Fi 6 & WPA2/WAP3 & Yes & TP-LINK & \CIRCLE & \CIRCLE & \CIRCLE 
        \\
        Archer C80 & Wi-Fi 5 & WPA2/WPA3 & Yes & TP-LINK & \CIRCLE & \Circle & \CIRCLE 
        \\
        Archer AX10 & Wi-Fi 6 & WPA2/WPA3 & Yes & TP-LINK & \CIRCLE & \CIRCLE & \CIRCLE 
        \\
        \midrule
        \rowcolor{mygray}
        AX3 & Wi-Fi 6 & WPA2/WPA3 & Yes & HUAWEI & \CIRCLE & \CIRCLE & \CIRCLE
        \\
        \rowcolor{mygray}
        WS7200 & Wi-Fi 6 & WPA2 & Yes & HUAWEI & \CIRCLE & \CIRCLE & \CIRCLE
        \\
        \rowcolor{mygray}
        WS7100 & Wi-Fi 6 & WPA2 & Yes & HUAWEI & \CIRCLE & \CIRCLE & \CIRCLE
        \\
        \rowcolor{mygray}
        WS318N & Wi-Fi 4 & WPA2 & Yes & HUAWEI & \CIRCLE & \Circle & \Circle
        \\
        \midrule
        RT-AC66U & Wi-Fi 5 & WPA2 & Yes & ASUS & \CIRCLE & \CIRCLE & \CIRCLE
        \\
        RT-AC68U & Wi-Fi 5 & WPA2 & Yes & ASUS & \CIRCLE & \CIRCLE & \CIRCLE
        \\
        RT-AX86U & Wi-Fi 6 & WPA2/WPA3 & Yes & ASUS & \CIRCLE & \CIRCLE & \CIRCLE
        \\
        RT-AX82U & Wi-Fi 6 & WPA2/WPA3 & Yes & ASUS & \CIRCLE & \CIRCLE & \CIRCLE
        \\
        \midrule
        \rowcolor{mygray}
        AC 6 & Wi-Fi 5 & WPA2 & Yes & Tenda & \CIRCLE & \Circle & \Circle 
        \\
        \rowcolor{mygray}
        AC 8 & Wi-Fi 5 & WPA2 & Yes & Tenda & \CIRCLE & \Circle & \CIRCLE 
        \\
        \rowcolor{mygray}
        AC 23 & Wi-Fi 5 & WPA2 & Yes & Tenda & \CIRCLE & \CIRCLE & \CIRCLE 
        \\
        \rowcolor{mygray}
        F9 & Wi-Fi 4 & WPA2 & No & Tenda & \Circle & \Circle & \CIRCLE 
        \\
        \midrule
        AX1800 & Wi-Fi 6 & WPA2/WPA3 & Yes & Netgear & \CIRCLE & \Circle & \CIRCLE 
        \\
        AX5400 & Wi-Fi 6 & WPA2/WPA3 & Yes & Netgear & \CIRCLE & \Circle & \CIRCLE 
        \\
        \midrule
        \rowcolor{mygray}
        E5600 & Wi-Fi 5 & WPA2 & Yes & Linksys & \CIRCLE & \CIRCLE & \CIRCLE 
        \\
        \rowcolor{mygray}
        E7350 & Wi-Fi 6 & WPA2/WPA3 & Yes & Linksys & \CIRCLE & \CIRCLE & \CIRCLE 
        \\
        \rowcolor{mygray}
        E8450 & Wi-Fi 6 & WPA2/WPA3 & Yes & Linksys & \CIRCLE & \Circle & \CIRCLE 
        \\
        \midrule
        RG-EW1200G PRO & Wi-Fi 5 & WPA2 & Yes & Ruijie & \Circle & \Circle & \CIRCLE 
        \\
        M32 & Wi-Fi 6 & WPA2 & Yes & Ruijie & \Circle & \Circle & \CIRCLE 
        \\
        \midrule
        \rowcolor{mygray}
        N21 & Wi-Fi 5 & WPA2 & No & H3C & \CIRCLE & \Circle & \CIRCLE 
        \\
        \rowcolor{mygray}
        NX15 & Wi-Fi 6 & WPA2/WPA3 & Yes & H3C & \CIRCLE & \Circle & \CIRCLE 
        \\
        \rowcolor{mygray}
        B6 & Wi-Fi 6 & WPA2/WPA3 & Yes & H3C & \CIRCLE & \CIRCLE & \CIRCLE 
        \\
    \bottomrule
    \end{tabular}
    \begin{tablenotes} 
        \item \Circle~indicates that the security mechanism is not supported by the router, while \CIRCLE~indicates that it is supported.
    \end{tablenotes} 
    \label{tab:router_full_list}
    \end{threeparttable}
\end{table*} 

\section{Real-world Attacks}
\label{sec:study}
To assess the impact of our attack, we conduct an extensive investigation on 30 popular wireless routers and 80 real-world Wi-Fi networks. We analyze 30 popular wireless routers and find that all the evaluated routers cannot protect the victim from our attack. Besides, we conduct SSH DoS attack and web hijack attack on the victim (\ie our device) in the real-world Wi-Fi networks following the experimental setup and procedure in Section~\ref{sec:imp}.
The results reveal that our attack is successful\footnote{We conduct 10 iterations of the SSH DoS attack and the web hijack attack on the victim. In this context, the ``successful'' means that these two attacks can be successfully executed at least once in the real-world Wi-Fi network.} against 75 (93.75\%) of the 80 assessed Wi-Fi networks.

\subsection{Analysis of AP Routers}
\label{test ap}

To protect data transmission in the shared wireless channels, Wi-Fi Alliance has introduced multiple security mechanisms, ranging from WEP to the state-of-the-art WPA3. Although many vendors have released wireless routers that support WPA3, the majority of real-world Wi-Fi networks still utilize the WPA2 security mechanism~\cite{lindroos2022covid}. In our empirical study, we find that out of the 30 tested wireless routers, 14 support WPA3, while the remaining 16 only support WPA2.

The early WEP used RC4 algorithm for data encryption, whereas WPA2 replaced them with the AES-CCMP algorithm. In the latest WPA3 standard, AES-GCMP is proposed to be used as the encryption method for WPA3 Enterprise mode.
However, none of these security mechanisms can prevent the encrypted frame size from leaking upper layer information. 
\zq{We evaluate the wireless routers based on the experimental setup and attack procedures described in Section~\ref{sec:imp}.}
Based on our empirical findings, we confirm that all 30 evaluated wireless routers could not protect the supplicant from our attacks. 

Table~\ref{tab:router_full_list} shows detailed information on 30 tested wireless routers in our investigation.
Take the first line for example, 
the evaluated router ``Mi 4C'' manufactured by Xiaomi belongs to the older Wi-Fi generation (Wi-Fi 4) and lacks support for IPv6 and WPA3.
As outlined in the product description, the ``Mi 4C'' device offers support for various security features. These include a built-in firewall that allows administrators to define packet forwarding rules, a flood defense mechanism that restricts malicious flood traffic to prevent DoS attacks, and MAC address filtering, which enables network access authorization based on hardware addresses. In our investigation, all tested routers claim to support different security mechanisms to prevent various attacks. However, our study demonstrates that the existing security mechanisms are inadequate against our attack.

\begin{table*}[htbp]\small
\centering
    \begin{threeparttable}
    \caption{Experimental results in 30 real-world Wi-Fi networks.}
    \begin{tabular}{c c c c c c c c c}
    \toprule  %
        \textbf{\tabincell{c}{No.}} &
        \textbf{\tabincell{c}{SSID}} &  \textbf{\tabincell{c}{AP Vendor}} & \textbf{\tabincell{c}{IPv4/IPv6}} & \textbf{\tabincell{c}{PHY model}} & \textbf{\tabincell{c}{AP isolation}} & \textbf{\tabincell{c}Wi-Fi channel} &\textbf{\tabincell{c}{SSH DoS}} & \textbf{\tabincell{c}{Web hijack}} 
        \\
        \midrule
        \rowcolor{mygray}
        1 & Bookstore 1 & ADSLR & \LEFTcircle & 802.11n/ac & No & 6, 161 & 7/10 & 6/10 
        \\
        \rowcolor{mygray}
        2 & Bookstore 2 & HUAWEI & \LEFTcircle & 802.11n/ac/ax & No & 11, 44 & 7/10 & 7/10 
        \\
        \rowcolor{mygray}
        3 & Bookstore 3 & Xiaomi & \LEFTcircle & 802.11n/ac & No & 6, 149 & 8/10 & 7/10 
        \\
        \midrule
        4 & Coffee Shop 1 & TP-LINK & \LEFTcircle & 802.11n/ac & No & 6, 60 & 8/10 & 6/10 
        \\
        5 & Coffee Shop 2 & Wimaster & \LEFTcircle & 802.11n/ac & Yes & 1, 48 & 7/10 & 6/10 
        \\
        6 & Coffee Shop 3 & Tenda & \CIRCLE & 802.11n/ac & No & 4, 153 & 6/10 & 5/10 
        \\
        \midrule
        \rowcolor{mygray}
        7 & Restaurant 1 & D-Link & \LEFTcircle & 802.11n/ac & No & 5, 149 & 7/10 & 5/10 
        \\
        \rowcolor{mygray}
        8 & Restaurant 2 & Ruijie & \LEFTcircle & 802.11n/ac & Yes & 11, 64 & 6/10 & 4/10 
        \\
        \rowcolor{mygray}
        9 & Restaurant 3 & iKuai & \LEFTcircle & 802.11n/ac & No & 1, 48 & 5/10 & 3/10 
        \\
        \midrule
        10 & Office building 1 & TP-LINK & \LEFTcircle & 802.11n/ac & No & 11, 36, 40 & 7/10 & 6/10
        \\
        11 & Office building 2 & H3C & \CIRCLE & 802.11n/ac & No & 1, 48, 153 & 8/10 & 7/10
        \\
        12 & Office building 3 & Netcore & \LEFTcircle & 802.11n/ac & Yes & 6, 149 & 8/10 & 6/10
        \\
        \midrule
        \rowcolor{mygray}
        13 & Enterprise 1 & TP-LINK & \LEFTcircle & 802.11n/ac & No & 6, 36 & 6/10 & 6/10 
        \\
        \rowcolor{mygray}
        14 & Enterprise 2 & HUAWEI & \LEFTcircle & 802.11n/ac & Yes & 11, 157 & 7/10 & 6/10 
        \\
        \rowcolor{mygray}
        15 & Enterprise 3 & Ruijie & \LEFTcircle & 802.11n/ac & Yes & 1, 11, 40, 149 & 6/10 & 5/10 
        \\
        \midrule
        16 & Fast Food Restaurant 1 & Wimaster & \LEFTcircle & 802.11n/ac/ax & No & 6, 161, 149 & 6/10 & 4/10 
        \\
        17 & Fast Food Restaurant 2 & TP-LINK & \LEFTcircle & 802.11n/ac & No & 3, 157 & 7/10 & 6/10 
        \\
        18 & Fast Food Restaurant 3 & Ruijie & \LEFTcircle & 802.11n/ac & No & 1, 44 & 6/10 & 6/10 
        \\
        \midrule
        \rowcolor{mygray}
        19 & Cinema 1 & HUAWEI & \LEFTcircle & 802.11n/ac & No & 1, 157 & 7/10 & 6/10 
        \\
        \rowcolor{mygray}
        20 & Cinema 2 & Ruijie & \LEFTcircle & 802.11n & No & 6 & 7/10 & 6/10 
        \\
        \rowcolor{mygray}
        21 & Cinema 3 & H3C & \LEFTcircle & 802.11n/ac & No & 10, 149 & 7/10 & 5/10 
        \\
        \midrule
        22 & Hotel 1 & HUAWEI & \LEFTcircle & 802.11n/ac & No & 6, 44 & 8/10 & 7/10 
        \\
        23 & Hotel 2 & D-Link & \LEFTcircle & 802.11n/ac & No & 1, 48 & 6/10 & 5/10 
        \\
        24 & Hotel 3 & Xiaomi & \LEFTcircle & 802.11n & Yes & 1 & 5/10 & 4/10 
        \\
        \midrule
        \rowcolor{mygray}
        25 & Experience Store 1 & HUAWEI & \LEFTcircle & 802.11n/ac & No & 1, 36 & 7/10 & 6/10 
        \\
        \rowcolor{mygray}
        26 & Experience Store 2 & HUAWEI & \LEFTcircle & 802.11n/ac & No & 11, 149 & 7/10 & 6/10 
        \\
        \rowcolor{mygray}
        27 & Experience Store 3 & Tenda & \LEFTcircle & 802.11n/ac & No & 4,153 & 6/10 & 5/10 
        \\
        \midrule
        28 &   Campus 1 & Xiaomi & \LEFTcircle & 802.11n/ac & No & 9, 36 & 6/10 & 4/10 
        \\
        29 &  Campus 2 & Ruijie & \LEFTcircle & 802.11n/ac & No & 1, 44 & 7/10 & 6/10
        \\
        30 &  Campus 3 & H3C & \LEFTcircle & 802.11n/ac & No & 1, 6, 40, 64  & 6/10 & 6/10
        \\
    \bottomrule
    \end{tabular}
    \begin{tablenotes} 
		\item \LEFTcircle ~means IPv4 only and \CIRCLE ~means both IPv4 and IPv6 are supported. 
    \end{tablenotes} 
    \label{tab:network}
    \end{threeparttable}
\end{table*}

\subsection{Real-world Wi-Fi Networks Evaluation}
\label{sec:real-world wifi}
The Wi-Fi scenarios we tested cover a wide range of public settings, including coffee shops, restaurants, hotels, cinemas, and bookstores. The experimental results illustrate that over 93\% (\ie 75 out of 80) of the evaluated Wi-Fi networks are vulnerable to our attack.
\zq{By exploiting the encrypted frame size side-channel, the attacker can conduct SSH DoS and web hijacking attacks in the real-world Wi-Fi networks, achieving average success rates of 63.73\% and 51.36\% respectively in our evaluation.}
Next, we elaborate on the evaluation results. 

\begin{figure}[htbp]
\centerline{\includegraphics[width=\linewidth]{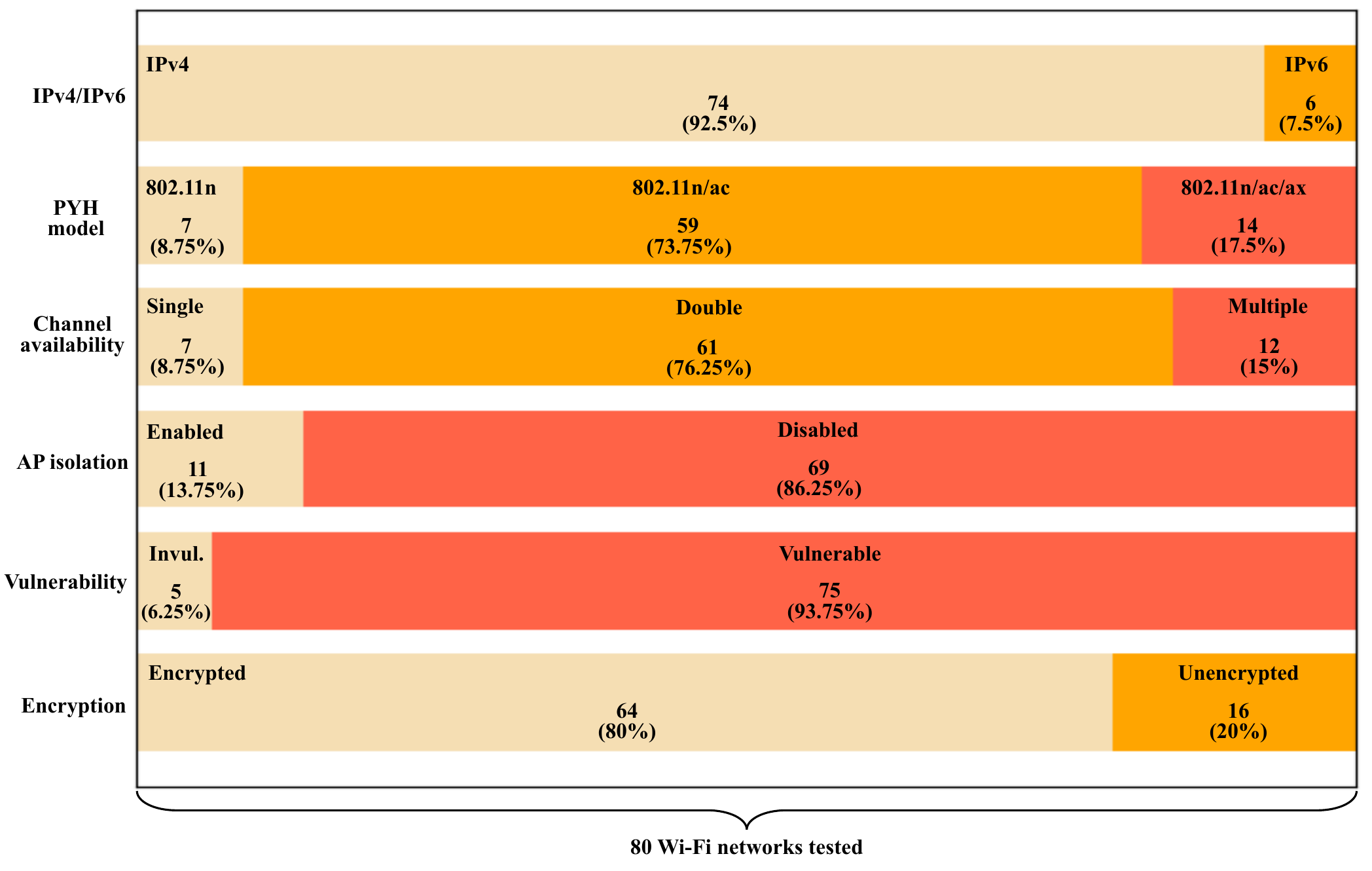}}
\caption{Attack evaluation on 80 real-world Wi-Fi networks.}
\label{fig:result}
\end{figure}

As shown in Figure~\ref{fig:result}, out of the 80 Wi-Fi networks we assessed, 74 are found to be IPv4-only networks, while the remaining 6 have IPv6 capabilities\footnote{Despite the accelerated deployment of IPv6 networks in recent years~\cite{Akamai}, the adoption of IPv6 support in Wi-Fi networks remains relatively limited.}. This can be attributed, in part, to the lack of IPv6 support in legacy wireless routers and the limited incentive for merchants to invest in new wireless routers.

The 802.11n/ac standards are predominantly utilized (73.75\%) in real-world Wi-Fi networks. This indicates that these Wi-Fi networks support two frequency bands (\ie 2.4 GHz and 5 GHz), with their physical layer models based on the 802.11n and 802.11ac standards, respectively.
There is only 17.5\% (14 out of 80) of the evaluated Wi-Fi networks support 802.11ax.
This is consistent with our expectations, as the 802.11ax standard was defined in 2019 and would require more time for widespread deployment.
Furthermore, the simultaneous utilization of two channels in Wi-Fi networks is the most prevalent case (76.25\%) due to the widespread support and default configuration of dual-channel capabilities in wireless routers. In certain scenarios, such as office buildings, Wi-Fi networks employ multiple wireless channels to enhance network performance. In our study, we identify 12 Wi-Fi networks that utilize multiple wireless channels. At first glance, the usage of multiple wireless channels might appear as a minor hurdle to our attack, as the attacker needs to perform additional channel scanning to determine the specific channel employed by the victim.
Indeed, the attacker can enhance the success rate by employing a channel ``eviction'' strategy, which will be explained in detail in Section~\ref{sec:discussion}. Additionally, we encounter several Wi-Fi networks that operate on a single channel. When questioned, network administrators cited security considerations as the rationale behind this choice, although they did not provide any further specifics.

Out of all the evaluated networks, 16 (20\%) of them are open and do not encrypt the supplicant's data frames. This means that an attacker can potentially access and view the contents of all supplicant's frames transmitted on these networks, representing a significant breach of supplicant privacy. 
The remaining 64 (80\%) Wi-Fi networks utilize WPA2/WPA3 to encrypt the supplicant's wireless frame.
It is worth noting that out of the 80 Wi-Fi networks evaluated, 11 (13.75\%) of them have AP isolation enabled. In these Wi-Fi networks, the attacker obtains the victim's MAC and IP address by leveraging the DHCP mechanism and injects malicious packets into the victim through the AP's external port\footnote{We do not encounter Wi-Fi networks with both AP isolation enabled and MFP required in our evaluations.}. However, our attack encounters failure in five Wi-Fi networks. Among them, one network is equipped with reverse path authentication~\cite{rfc2827, rfc8704}, preventing the attacker from sending packets from the WLAN to the AP's external port. In the remaining four Wi-Fi networks, the attacker cannot obtain the AP's external port as the gateway does not respond to ICMP ping messages.

\zq{We elaborate on the experimental results of 30 encrypted Wi-Fi networks in Table~\ref{tab:network}. Details of all 80 evaluated Wi-Fi networks are presented in Appendix~\ref{sec:appendix}.}
We take the first row of Table~\ref{tab:network} as an example to analyze the results. In our study, the SSID ``Bookstore 1'' indicates a Wi-Fi network that is accessible in a bookstore. 
It is common practice to set the Wi-Fi SSID as the organization name, which may expose the organization's identity. Therefore, to protect anonymity, we have anonymized the Wi-Fi SSID in this paper.
This bookstore's Wi-Fi network only supports IPv4 and does not have AP isolation enabled, while its AP is produced by ADSLR. This AP provides two access channels (\ie 6 and 161) and employs the 802.11n and 802.11ac standards.
The TCP connection of the victim supplicant can be hijacked using the attack presented in Section~\ref{sec:4}. 
The success rates for conducting SSH DoS and web hijacking on the victim supplicant are 70\% and 60\%, respectively.

Within the evaluated vulnerable Wi-Fi networks, we have observed a range of success rates for our attack, varying from 30\% to 80\%. 
The principal factor influencing this variance is the heterogeneous wireless environments (\eg different wireless interference and channel contention) encountered in real-world Wi-Fi networks, leading to varying capabilities for attacker to capture the victim's Wi-Fi frames. Factors such as wireless interference (\eg from microwave ovens and Bluetooth devices) and channel contention can hinder the attacker's ability to capture the victim's Wi-Fi frames,  resulting in the failure of attacks. For instance, as illustrated in Table~\ref{tab:network}, Coffee Shop 1, situated on a university campus, experiences less wireless interference and channel contention compared to Restaurant 3, located within a large shopping mall. Consequently, the success rate of attacks in the latter Wi-Fi network is lower due to the elevated interference and contention in that environment. We will conduct a more comprehensive analysis of the factors influencing the success of our attack in Section~\ref{sec:discussion}.

\section{Discussion}
\label{sec:discussion}
Our attack relies on the observation of the victim's encrypted frame size. However, the attacker's ability to monitor the victim's frames may be hindered by wireless interference, Wi-Fi channel contention and frame aggregation.
These factors can directly influence the success and effectiveness of our attack. We delve into the details of these factors in this section.

\parab{Wireless Interference.} 
The transmission of Wi-Fi frames over the wireless medium is susceptible to losses. These losses are often a result of interference, leading to a diminished signal to interference and noise ratio (SINR) at the receiver~\cite{zhang2012frame}. A low SINR decreases the likelihood of successfully decoding all the bits in the frame. Wi-Fi networks face various sources of interference, including microwave ovens, Bluetooth devices, radar signals, and more. Consequently, frame reception failures are frequent occurrences in Wi-Fi networks~\cite{khan2015smart}.
Due to wireless interference, the attacker may not be able to sniff all of the victim's encrypted frames.
To mitigate wireless interference, we employ a straightforward yet efficient multiple verification strategy. This strategy involves using multiple monitoring wireless network interface cards and performing repeated verifications of the inferred values. By leveraging multiple wireless network interface cards and verifying the inferred values multiple times, we increase the reliability and accuracy of our analysis.

\parab{Channel Contention.} APs and supplicants based on the 802.11 standards use Carrier Sense Multiple Access with Collision Avoidance (CSMA/CA) to compete equally for the occupation of the wireless channel. 
Before transmitting frames, wireless channel listening is conducted to ensure that the channel is not occupied. Frames are transmitted only after verifying the channel's availability.
Due to channel contention, there may be an uncertain delay or even frame dropping in the victim's responses to the probe packets. 
This uncertain response delay or frame dropping is the primary reason for the fluctuating success rate of our attack because the attacker needs to analyze the victim's encrypted frames within a time slice after the probe packets are sent.
To mitigate channel contention, we propose a channel ``eviction'' strategy. The attacker can evict other supplicants from the channel used by the victim.
Specifically, the attacker impersonates the AP and sends decertification frames to the supplicant, causing it to detach from the current channel of the AP\footnote{In the Wi-Fi network with MFP (Management Frame Protection) enabled, attackers can exploit implementation vulnerabilities to force the supplicant to detach from the current channel~\cite{DBLP:conf/wisec/SchepersRV22, schepersframing}.}. The supplicant will attempt to reconnect to the Wi-Fi network, but after encountering several disconnections, it will switch to another channel.
This strategy requires the Wi-Fi network to support multiple access channels. Fortunately, most Wi-Fi networks provide more than one access channel, as shown in Section~\ref{sec:real-world wifi}. 
Note that the channel switching (\ie our ``eviction'' strategy to cause other supplicants to detach from the current channel) is transparent to the users. The only impact is that the user may experience a brief (a few seconds) network jitter during the channel switching. 

\parab{Frame Aggregation.} The MAC layer frame aggregation technique is proposed in the 802.11n standard~\cite{5307322} to improve the throughput and efficiency of WLANs by combining multiple data packets into a single transmission unit. There are two methods available to perform frame aggregation, \ie aggregate MAC protocol service unit (A-MSDU) and aggregate MAC protocol data unit (A-MPDU). 
The main difference between MSDU and MPDU is that the latter has a MAC header through 802.11 protocol encapsulation while the former becomes MPDU after adding integrity check MIC, encryption, sequence number assignment, CRC checksum, and MAC header.
The A-MPDU has no impact on our attack because each MPDU has a complete MAC header and the attacker can distinguish the encryption payload size of each MPDU.
However, if the victim triggers A-MSDU, multiple packets will be encrypted together, preventing the attacker from inferring TCP information based on the encrypted frame size. 
Fortunately, the server's responses triggered by attackers are rarely aggregated into A-MSDUs.
In the following, we analyze the reasons why A-MSDU frames are not triggered.

The A-MSDU completes when the size of the waiting packets reaches the maximum A-MSDU threshold or the maximum delay of the oldest packets reaches a pre-assigned value. Its maximum size can be 3839 or 7935 bytes, depending on the throughput capacity of the station (STA). 
The size can be found from the High-Throughput (HT) capabilities element of the HT STA release.
In case the aggregated frame size does not reach the aggregation threshold, the MSDU buffer queue waits for new MSDUs to reach the MAC layer.
But if the maximum delay exceeds the preset maximum, the aggregated frame will be immediately inserted into the channel, even if the aggregated frame size does not reach the aggregation threshold.
The maximum delay is typically set to 1 \textmu s ~\cite{skordoulis2008ieee}. 
Additionally, only frames with the same receiver and the same Traffic Identifier (TID) can be aggregated together using A-MSDU.
In our attack, few A-MSDU frames are observed.
We speculate that this absence may be due to the attacker sending probe packets at a low rate (compared to 1 \textmu s)\footnote{The attacker can control the time interval between sending probe packets to be greater than 1 \textmu s.}, and the response packets having a different TID than the background traffic (\eg video packets).
Consequently, the TCP responses triggered by the probe packets are not aggregated into A-MSDU.

\section{Countermeasure}
The root cause of our attack can be attributed to the combination of two specific conditions. The first condition is the inconsistent response of the TCP stack under different trigger conditions. The second condition is the leakage of TCP connection information through the frame size side channel.
As a result, we propose two countermeasures to mitigate this vulnerability, one derived from the 802.11 standard and the other from the TCP stacks.

\label{sec:countermeasure}
\noindent\textbf{Defenses in 802.11 Standard.} As Wi-Fi networks rely on shared wireless media, any 802.11-compliant device has the capability to sniff all Wi-Fi frames. To maintain the confidentiality and integrity of Wi-Fi frames, encryption mechanisms are commonly employed in Wi-Fi networks. Although Wi-Fi networks encrypt their frames transmitted in the wireless channel, there exists a strong correlation between the encrypted frame size and the upper layer applications. This correlation allows an off-path attacker to analyze the encrypted frame size, infer the victim's TCP information, and subsequently conduct the TCP hijacking attack.
Adjusting the security mechanisms of the 802.11 standards so that the AP or supplicant dynamically pads the size of encrypted frames is one possible countermeasure.
This countermeasure may require changes and redesign at the Wi-Fi standard level. We are currently in discussions with the Wi-Fi Alliance regarding this countermeasure.

\noindent\textbf{Defenes in TCP Stacks.} 
The packet validation logic in the latest TCP specification handles valid and invalid incoming packets differently depending on whether a response needs to be generated and the type of response required. This difference is reflected in two aspects: (\rmnum{1}) The number of response packets is different. For example, 
during the verification of the acknowledgment number,
one challenge \ack will be triggered if the packet's acknowledgment number falls within the challenge window. If it falls outside the window, no packet will be sent. 
(\rmnum{2}) The response packets have different types. 
The type of TCP packet can be identified by its size, which is influenced by the varying header options.
For instance, the size of a \sack packet is 78 bytes, whereas a \rst packet is only 54 bytes in size.
An attacker can infer the state of a TCP connection by observing the size of the response packets, which are encrypted frames in our attack.
To resolve this problem, a possible solution is to revise the TCP specification by obfuscating the header sizes for different types of TCP packets (\eg \rst, \ack, and \sack) and adjusting the trigger conditions for the challenge \ack.

\section{Related Work}
\label{sec:9}
\noindent\textbf{Traffic Analysis.} The prior traffic analysis works endeavors aimed to analyze users' encrypted traffic and compromise their privacy by, for instance, tracking the applications~\cite{shen2019encrypted, petagna2019peel, shen2021accurate, van2020flowprint, bahramali2020practical} and websites~\cite{rimmer2017automated, shen2020fine, wang2014effective, hayes2016k} they accessed.  Ede \et designed a semi-supervised scheme for creating application fingerprints from encrypted network traffic of mobile devices~\cite{van2020flowprint}. Shen \et used Graph Neural Networks to identify decentralized applications from encrypted traffic~\cite{shen2021accurate}.
Hayes \et established that website fingerprinting attacks are a serious threat to online privacy~\cite{hayes2016k}.
Rimmer \et harnessed deep learning for web fingerprinting, which de-anonymizes Tor traffic by classifying encrypted web traffic~\cite{rimmer2017automated}.
Furthermore, several academic studies delve into the privacy challenges associated with encrypted DNS~\cite{shulman2014pretty, siby2019encrypted, houser2019investigation, bushart2020padding}. Shulman proposed that encryption alone may not be sufficient to protect users~\cite{shulman2014pretty}, and Siby \et demonstrated that classifying encrypted DNS traffic can jeopardize the user privacy~\cite{siby2019encrypted}.

Our attack and prior research on traffic analysis both involve extracting information from encrypted packets. Nonetheless, there are three key distinctions between our attack and traffic analysis work. Firstly, while previous work relies on an on-path attack model, our attack does not require such positioning. Secondly, traffic analysis typically involves the creation of a database as a prerequisite, whereas our attack operates without this necessity. Finally, existing traffic analysis work focuses on upper-layer applications, while our attack interferes with the underlying transport protocol.

\noindent\textbf{Wi-Fi Attacks.}
While Wi-Fi serves as a widely used access method for end-users to connect to the internet, it presents higher security risks compared to wired LANs, such as Ethernet.
Public Wi-Fi networks, in particular, are susceptible to attacks due to their open-access nature. 
To safeguard wireless users in Wi-Fi networks, numerous security mechanisms have been proposed in recent years, including WEP, WPA, WPA2, and WPA3~\cite{wifi}. Nevertheless, existing researches~\cite{tews2009practical, fluhrer2001weaknesses, ohigashi2009practical, vanhoef2017key, vanhoef2018release, vanhoef2020dragonblood, vanhoef2021fragment} have revealed implementation vulnerabilities or design flaws in these security mechanisms that can compromise Wi-Fi networks. For example, WPA is vulnerable to key recovery attacks~\cite{tews2009practical, ohigashi2009practical} and dictionary attacks~\cite{moskowitz2003weakness}.
Subsequently, WPA2 and WPA3 were introduced to mitigate these vulnerabilities.
However, recent research indicates that WPA2 is susceptible to KRACK attacks~\cite{vanhoef2017key, vanhoef2018release}, and WPA3 can be compromised by downgrade or dictionary attacks~\cite{vanhoef2020dragonblood}. 
Besides, recent studies~\cite{vanhoef2021fragment, schepersframing} have revealed that attackers can leverage the design flaws of Wi-Fi networks to circumvent these security mechanisms. Unlike the aforementioned studies, our attack does not require cracking or circumventing these security mechanisms.

In addition to Wi-Fi network cracking, extensive research has been conducted on traffic hijacking within Wi-Fi networks.
Attackers can execute an Evil Twins attack by deploying a rogue AP to hijack the traffic of victim supplicants~\cite{gao2021nationwide, han2011timing, alsahlany2018risk, orsi2019understanding}. 
Additionally, rogue DHCP and ARP poisoning are recognized as common threats in Wi-Fi networks. Notably, these attacks have been subject to extensive research, leading to the development of countermeasures, including rogue AP and rogue DHCP detection~\cite{Linksys-rogueAP, Huawei-rogueAP, agarwal2017discrete, DHCP-Sentry} and ARP protection~\cite{ARP-AntiSpoofer, shARP, 360-ARP}.
Recently, Feng \et revealed vulnerabilities in the implementation of IP source address checking in wireless routers, enabling attackers to hijack victim's traffic in Wi-Fi networks using ICMP redirect messages~\cite{feng2022man}. 
Yang \et proposed exploiting flaws in \rst packet inspection implementation in wireless routers to manipulate NAT mapping states and hijack the victim's TCP connection~\cite{yangexploiting}.
In contrast, our attack does not rely on such implementation flaws in wireless routers. Instead, it reveals a novel fundamental security vulnerability in the 802.11 standards, affecting all Wi-Fi networks.

\noindent\textbf{Side Channel Attacks.}
In many cases, off-path attackers rely on a side channel to carry out their attacks, where blind attackers can extract significant information from this channel~\cite{ensafi2010idle, ensafi2014detecting, zhang2015original, alexander2015off, gilad2012spying, qian2010investigation, pearce2017augur, zhang2016high}. In one instance, Ensafi \et utilized the side channel of global IPID~\cite{ensafi2010idle} counters to perform idle port scans and network protocol analyses. They also proposed that these counters could be leveraged to detect intentional packet drops.
In another example, Alexander \et inferred the round-trip time (RTT) between two arbitrary hosts by examining the shared \syn backlog~\cite{alexander2015off}. 

In TCP connection hijacking attacks, side channels serve as potent tools for attackers. The IPID, in particular, has been a frequent target for exploitation. For example, Jeffrey \et utilized per-destination IPID counters to estimate the number of packets transmitted between two machines and even detect the presence of a TCP connection~\cite{knockel2014counting}. Similarly, Alexander \et used the IPID of triggered RST packets to identify the existence of the victim's TCP connection~\cite{alexander2019detecting}. In a recent instance, Feng \et manipulated the IPID assignment using ICMP to hijack the victim's TCP connection~\cite{feng2020off}. 
Besides IPID, the challenge \ack mechanism is another side channel exploited by off-path attackers.
Cao \et, for example, utilized the global rate limit of challenge \ack to infer and hijack TCP connections~\cite{cao2016off, DBLP:journals/ton/CaoQWDKM18}. Moreover, a timing side channel has been found in half-duplex Wi-Fi technology, which can be exploited by off-path attackers to inject data into the victim's TCP connection~\cite{chen2018off}. 
\zq{However, this method typically requires attackers to install a puppet on the victim client, which is not necessary for our attack.}
Tolley \et recently proposed a blind in/on-path attack in VPNs, aiming to infer the existence of, interfere with, or inject data into TCP connections forwarded through encrypted VPN tunnels~\cite{tolley2021blind}. Different from previous research, our work reveals a new side channel, \ie information leakage due to the encrypted frame size in Wi-Fi networks. This side channel can be exploited by a pure off-path attacker to hijack victim's TCP connections.

\section{Conclusion}
\label{sec:10}
In this paper, we present a new off-path TCP hijacking attack that takes advantage of the encrypted frame size in Wi-Fi networks to detect and hijack TCP connections belonging to a victim supplicant.
\zq{Our attack focuses on TCP connection hijacking, but non-TCP sessions may also be affected due to Wi-Fi layer information leakage, which we will explore in future work.} 
This side channel (\ie the observable frame size) vulnerability is an inherent flaw in Wi-Fi networks, specifically the 802.11 standards.
To execute our attack, attackers initially scan the WLAN to identify active victim supplicant, then analyze the victim's encrypted frame size to infer the 4-tuple, exact sequence number, and acceptable acknowledgment number of the victim's TCP connection. Specifically, our attacker can hijack the victim's TCP connection within 28 seconds.
We carry out our attack in typical Wi-Fi scenarios, and our evaluation demonstrates that this new off-path TCP hijacking attack can result in significant damage to upper-layer applications, such as SSH DoS and the injection of malicious data into web traffic. Moreover, we conduct comprehensive studies involving 80 real-world Wi-Fi networks and 30 popular wireless routers.
The results reveal that a majority of assessed Wi-Fi networks (75 out of 80) are vulnerable to our attack, and all tested routers fail to resist our attack. We have responsibly disclosed this vulnerability. While eliminating the side channel of encrypted frame size in Wi-Fi networks presents challenges, we propose several potential countermeasures to mitigate this vulnerability.

\section*{Acknowledgment}
We thank the anonymous reviewers for their thoughtful comments. This work was in part supported by
NSFC under U22B2025, Jiangsu Provincial Scientific Research Center of Applied Mathematics under Grant No. BK20233002, 
the National Science Foundation for Distinguished Young Scholars of China under No. 62425201, 
the Science Fund for Creative Research Groups of the National Natural Science Foundation of China under No. 62221003, 
the Key Program of the National Natural Science Foundation of China under No. 61932016 and No. 62132011. 
Ziqiang Wang, Ke Xu and Jianping Wu are corresponding authors.



%



\bibliographystyle{IEEEtranS}
\bibliography{reference}


\appendix

\section{The 80 evaluated Wi-Fi networks}
\label{sec:appendix}

All the 80 tested Wi-Fi networks are shown in Table~\ref{tab:all_network}.

\begin{table*}[htbp]\small
\renewcommand\arraystretch{0.1}
\centering
\scalebox{0.9}{
    \begin{threeparttable}
    \caption{Experimental results in 80 real-world Wi-Fi networks.}
    \begin{tabular}{c c c c c c c c c}
    \toprule  %
        \textbf{\tabincell{c}{No.}} &
        \textbf{\tabincell{c}{SSID}} &  \textbf{\tabincell{c}{AP Vendor}} & \textbf{\tabincell{c}{IPv4/IPv6}} & \textbf{\tabincell{c}{PHY model}} & \textbf{\tabincell{c}{AP isolation}} & \textbf{\tabincell{c}Wi-Fi channel} &\textbf{\tabincell{c}{SSH DoS}} & \textbf{\tabincell{c}{Web hijack}} 
        \\
        \midrule
        \rowcolor{mygray}
        1 & Bookstore 1 & ADSLR & \LEFTcircle & 802.11n/ac & No & 6, 161 & 7/10 & 6/10 
        \\
        \rowcolor{mygray}
        2 & Bookstore 2 & HUAWEI & \LEFTcircle & 802.11n/ac/ax & No & 11, 44 & 7/10 & 7/10 
        \\
        \rowcolor{mygray}
        3 & Bookstore 3 & UTT & \LEFTcircle & 802.11n & No & 1 & \usym{1F5F8} & \usym{1F5F8} 
        \\
        \rowcolor{mygray}
        4 & Bookstore 4 & Xiaomi & \LEFTcircle & 802.11n/ac & No & 6, 149 & 8/10 & 7/10 
        \\
        \rowcolor{mygray}
        5 & Bookstore 5 & TP-LINK & \LEFTcircle & 802.11n/ac & No & 7, 36 & 7/10 & 5/10 
        \\
        \rowcolor{mygray}
        6 & Bookstore 6 & Tenda & \LEFTcircle & 802.11n/ac & No & 9, 48 & 5/10 & 3/10 
        \\
        \rowcolor{mygray}
        7 & Bookstore 7 & Ruijie & \LEFTcircle & 802.11n/ac & No & 5, 149 & 6/10 & 5/10 
        \\
        \midrule
        8 & Coffee Shop 1 & TP-LINK & \LEFTcircle & 802.11n/ac & No & 6, 60 & 8/10 & 6/10 
        \\
        9 & Coffee Shop 2 & HUAWEI & \LEFTcircle & 802.11n/ac & No & 1, 36 & 8/10 & 7/10 
        \\
        10 & Coffee Shop 3 & Wimaster & \LEFTcircle & 802.11n/ac & Yes & 1, 48 & 7/10 & 6/10 
        \\
        11 & Coffee Shop 4 & Tenda & \CIRCLE & 802.11n/ac & No & 4, 153 & 6/10 & 5/10 
        \\
        12 & Coffee Shop 5 & TP-LINK & \LEFTcircle & 802.11n/ac & No & 153 & 7/10 & 5/10 
        \\
        13 & Coffee Shop 6 & Ruckus & \LEFTcircle & 802.11n/ac & No & 11, 157 & \usym{1F5F8} & \usym{1F5F8}
        \\
        14 & Coffee Shop 7 & Xiaomi & \LEFTcircle & 802.11n/ac & No & 8, 36 & 6/10 & 6/10
        \\
        15 & Coffee Shop 8 & HUAWEI & \LEFTcircle & 802.11n/ac/ax & No & 6, 48 & \usym{1F5F8} & \usym{1F5F8}
        \\
        \midrule
        \rowcolor{mygray}
        16 & Restaurant 1 & D-Link & \LEFTcircle & 802.11n/ac & No & 5, 149 & 7/10 & 5/10 
        \\
        \rowcolor{mygray}
        17 & Restaurant 2 & TP-LINK & \LEFTcircle & 802.11n/ac & No & 1, 153 & 6/10 & 6/10 
        \\
        \rowcolor{mygray}
        18 & Restaurant 3 & Ruijie & \LEFTcircle & 802.11n/ac & Yes & 11, 64 & 6/10 & 4/10 
        \\
        \rowcolor{mygray}
        19 & Restaurant 4 & iKuai & \LEFTcircle & 802.11n/ac & No & 1, 48 & 5/10 & 3/10 
        \\
        \rowcolor{mygray}
        20 & Restaurant 5 & TP-LINK & \LEFTcircle & 802.11n/ac & No & 2, 64  & \usym{1F5F8} & \usym{1F5F8} 
        \\
        \rowcolor{mygray}
        21 & Restaurant 6 & Xiaomi & \LEFTcircle & 802.11n/ac & No & 36 & 6/10 & 5/10 
        \\
        \rowcolor{mygray}
        22 & Restaurant 7 & ASUS & \CIRCLE & 802.11n/ac/ax & Yes & 3, 161  & \usym{2717} & \usym{2717} 
        \\
        \rowcolor{mygray}
        23 & Restaurant 8 & HUAWEI & \LEFTcircle & 802.11n/ac & No & 11, 157  & 5/10 & 4/10 
        \\
        \midrule
        24 & Office building 1 & TP-LINK & \LEFTcircle & 802.11n/ac & No & 11, 36, 40 & 7/10 & 6/10
        \\
        25 & Office building 2 & H3C & \CIRCLE & 802.11n/ac & No & 1, 48, 153 & 8/10 & 7/10
        \\
        26 & Office building 3 & Netcore & \LEFTcircle & 802.11n/ac & Yes & 6, 149 & 8/10 & 6/10
        \\
        27 & Office building 4 & ZTE & \LEFTcircle & 802.11n/ac & No & 11, 60 & 6/10 & 6/10
        \\
        28 & Office building 5 & H3C & \LEFTcircle & 802.11n/ac/ax & No & 11, 36, 52, 149 & \usym{1F5F8} & \usym{1F5F8}
        \\
        29 & Office building 6 & Linksys & \LEFTcircle & 802.11n/ac/ax & No & 11, 48 & 6/10 & 4/10
        \\
        30 & Office building 7 & HUAWEI & \LEFTcircle & 802.11n/ac & No & 9, 161 & 7/10 & 5/10
        \\
        31 & Office building 8 & TP-LINK & \LEFTcircle & 802.11n/ac & No & 1, 157 & 7/10 & 6/10
        \\
        \midrule
        \rowcolor{mygray}
        32 & Enterprise 1 & TP-LINK & \LEFTcircle & 802.11n/ac & No & 6, 36 & 6/10 & 6/10 
        \\
        \rowcolor{mygray}
        33 & Enterprise 2 & HUAWEI & \LEFTcircle & 802.11n/ac & Yes & 11, 157 & 7/10 & 6/10 
        \\
        \rowcolor{mygray}
        34 & Enterprise 3 & Ruijie & \LEFTcircle & 802.11n/ac & Yes & 1, 11, 40, 149 & 6/10 & 5/10 
        \\
        \rowcolor{mygray}
        35 & Enterprise 4 & H3C & \LEFTcircle & 802.11n/ac/ax & No & 10, 149 & \usym{1F5F8} & \usym{1F5F8} 
        \\
        \rowcolor{mygray}
        36 & Enterprise 5 & HUAWEI & \LEFTcircle & 802.11n/ac/ax & Yes & 6, 48, 161 & \usym{2717} & \usym{2717} 
        \\
        \rowcolor{mygray}
        37 & Enterprise 6 & PHICOMM & \LEFTcircle & 802.11n/ac & No & 6, 36 & 6/10 & 4/10 
        \\
        \rowcolor{mygray}
        38 & Enterprise 7 & H3C & \CIRCLE & 802.11n/ac & Yes & 1, 6, 64 & \usym{2717} & \usym{2717} 
        \\
        \midrule
        39 & Fast Food Restaurant 1 & Wimaster & \LEFTcircle & 802.11n/ac/ax & No & 6, 161, 149 & 6/10 & 4/10 
        \\
        40 & Fast Food Restaurant 2 & TP-LINK & \LEFTcircle & 802.11n/ac & No & 3, 157 & 7/10 & 6/10 
        \\
        41 & Fast Food Restaurant 3 & Wimaster & \LEFTcircle & 802.11n/ac & No & 11, 157 & \usym{1F5F8} & \usym{1F5F8} 
        \\
        42 & Fast Food Restaurant 4 & Ruijie & \LEFTcircle & 802.11n/ac & No & 1, 44 & 6/10 & 6/10 
        \\
        43 & Fast Food Restaurant 5 & HUAWEI & \LEFTcircle & 802.11n/ac/ax & No & 6, 153 & 5/10 & 4/10
        \\
        44 & Fast Food Restaurant 6 & Tenda & \LEFTcircle & 802.11n/ac & No & 11, 60 & \usym{1F5F8} & \usym{1F5F8}
        \\
        45 & Fast Food Restaurant 7 & Xiaomi & \LEFTcircle & 802.11n/ac & No & 1, 52 & 6/10 & 5/10
        \\
        46 & Fast Food Restaurant 8 & ZTE & \LEFTcircle & 802.11n/ac & No & 11, 40 & 5/10 & 3/10
        \\
        \midrule
        \rowcolor{mygray}
        47 & Cinema 1 & HUAWEI & \LEFTcircle & 802.11n/ac & No & 1, 157 & 7/10 & 6/10 
        \\
        \rowcolor{mygray}
        48 & Cinema 2 & WayOS & \LEFTcircle & 802.11n/ac & No & 11, 157 & \usym{1F5F8} & \usym{1F5F8}
        \\
        \rowcolor{mygray}
        49 & Cinema 3 & Ruijie & \LEFTcircle & 802.11n & No & 6 & 7/10 & 6/10 
        \\
        \rowcolor{mygray}
        50 & Cinema 4 & H3C & \LEFTcircle & 802.11n/ac & No & 10, 149 & 7/10 & 5/10 
        \\
        \rowcolor{mygray}
        51 & Cinema 5 & HUAWEI & \LEFTcircle & 802.11n/ac & No & 3, 161 & \usym{1F5F8} & \usym{1F5F8} 
        \\
        \midrule
        52 & Hotel 1 & HUAWEI & \LEFTcircle & 802.11n/ac & No & 6, 44 & 8/10 & 7/10 
        \\
        53 & Hotel 2 & Ruijie & \LEFTcircle & 802.11n & No & 1, 11 & 8/10 & 6/10 
        \\
        54 & Hotel 3 & D-Link & \LEFTcircle & 802.11n/ac & No & 1, 48 & 6/10 & 5/10 
        \\
        55 & Hotel 4 & Xiaomi & \LEFTcircle & 802.11n & Yes & 1 & 5/10 & 4/10 
        \\
        56 & Hotel 5 & TP-LINK & \LEFTcircle & 802.11n/ac & No & 9, 48 & 6/10 & 5/10 
        \\
        57 & Hotel 6 & China Unicom & \LEFTcircle & 802.11n/ac & Yes & 1, 11, 36, 157 & \usym{2717} & \usym{2717} 
        \\
        58 & Hotel 7 & HUAWEI & \LEFTcircle & 802.11n/ac & No & 60 & 6/10 & 4/10 
        \\
        59 & Hotel 8 & Wimaster & \LEFTcircle & 802.11n/ac/ax & No & 6, 56 & 7/10 & 5/10 
        \\
        \midrule
        \rowcolor{mygray}
        60 & Experience Store 1 & HUAWEI & \LEFTcircle & 802.11n/ac & No & 1, 36 & 7/10 & 6/10 
        \\
        \rowcolor{mygray}
        61 & Experience Store 2 & HUAWEI & \LEFTcircle & 802.11n/ac & No & 11, 149 & 7/10 & 6/10 
        \\
        \rowcolor{mygray}
        62 & Experience Store 3 & Tenda & \LEFTcircle & 802.11n/ac & No & 4,153 & 6/10 & 5/10 
        \\
        \rowcolor{mygray}
        63 & Experience Store 4 & TP-LINK & \LEFTcircle & 802.11n/ac & No & 11, 36 & 5/10 & 3/10 
        \\
        \rowcolor{mygray}
        64 & Experience Store 5 & Xiaomi & \LEFTcircle & 802.11n/ac & Yes & 6, 64 & \usym{2717} & \usym{2717} 
        \\
        \rowcolor{mygray}
        65 & Experience Store 6 & H3C & \LEFTcircle & 802.11n/ac/ax & No & 8, 52 & 5/10 & 4/10 
        \\
        \rowcolor{mygray}
        66 & Experience Store 7 & Ruckus & \LEFTcircle & 802.11n/ac & No & 1, 56 & \usym{1F5F8} & \usym{1F5F8} 
        \\
        \midrule
        67 &   Campus 1 & Xiaomi & \LEFTcircle & 802.11n/ac & No & 9, 36 & 6/10 & 4/10 
        \\
        68 &  Campus 2 & Ruijie & \LEFTcircle & 802.11n/ac & No & 1, 44 & 7/10 & 6/10
        \\
        69 &  Campus 3 & H3C & \LEFTcircle & 802.11n/ac & No & 1, 6, 40, 64  & 6/10 & 6/10
        \\
        70 &  Campus 4 & ASUS & \LEFTcircle & 802.11n/ac/ax & No & 6, 40  & 5/10 & 3/10
        \\
        71 &  Campus 5 & H3C & \CIRCLE & 802.11n/ac & No & 1, 6, 36  & \usym{1F5F8} & \usym{1F5F8}
        \\
        72 &  Campus 6 & Netgear & \LEFTcircle & 802.11n/ac/ax & No & 3, 149  & 6/10 & 5/10
        \\
        73 &  Campus 7 & H3C & \CIRCLE & 802.11n/ac & No & 11, 48 & \usym{1F5F8} & \usym{1F5F8}
        \\
        \midrule
        \rowcolor{mygray}
        74 & Shopping Mall 1 & Ruijie & \LEFTcircle & 802.11n/ac & No & 11, 149 & 4/10 & 3/10 
        \\
        \rowcolor{mygray}
        75 & Shopping Mall 2 & Ruckus & \LEFTcircle & 802.11n & No & 1, 149 & \usym{1F5F8} & \usym{1F5F8} 
        \\
        \rowcolor{mygray}
        76 & Shopping Mall 3 & HUAWEI & \LEFTcircle & 802.11n/ac & No & 1, 157 & 6/10 & 6/10 
        \\
        \rowcolor{mygray}
        77 & Shopping Mall 4 & SUNDRAY & \LEFTcircle & 802.11n & No & 6, 11 & \usym{1F5F8} & \usym{1F5F8} 
        \\
        \rowcolor{mygray}
        78 & Shopping Mall 5 & HUAWEI & \LEFTcircle & 802.11n/ac & No & 1, 11, 48, 157 & 7/10 & 5/10 
        \\
        \rowcolor{mygray}
        79 & Shopping Mall 6 & H3C & \LEFTcircle & 802.11n/ac/ax & No & 1, 36, 44 & 6/10 & 4/10 
        \\
        \rowcolor{mygray}
        80 & Shopping Mall 7 & TP-LINK & \LEFTcircle & 802.11n & No & 11 & \usym{1F5F8} & \usym{1F5F8} 
        \\
    \bottomrule
    \end{tabular}
    \begin{tablenotes} 
		\item \LEFTcircle ~means IPv4 only and \CIRCLE ~means both IPv4 and IPv6 are supported.
        \item \usym{1F5F8} indicates the WiFi network does not encrypt frames, letting the attacker obtain the victim's TCP connection information directly.
        \item \usym{2717} indicates the attack failed in the Wi-Fi network.
    \end{tablenotes} 
    \label{tab:all_network}
    \end{threeparttable}
    }
\end{table*}

\end{document}